\documentclass{raa_twocolumn}

\usepackage{graphicx,times}
\usepackage{natbib}
\usepackage{amssymb,amsmath}
\usepackage{booktabs}
\usepackage{multirow}
\usepackage{xcolor}
\bibpunct{(}{)}{;}{a}{}{,}
\usepackage[
  pagebackref=true,
  colorlinks=true,
  linkcolor=blue,
  citecolor=blue,
  urlcolor=blue,
  filecolor=blue
]{hyperref}

\graphicspath{{./}}

\begin{document}

\title{Gamma-Ray Emission from the Crab Pulsar: A 17-Year \textit{Fermi}-LAT Reanalysis}

\volnopage{Vol.0 (2026) No.0, 000--000}
\setcounter{page}{1}

\author{Liancheng Zhou \inst{1,2}
\and Yunlu Gong \inst{1}
\and Jun Fang \inst{1}
\and Li Zhang \inst{1}}

\institute{Department of Astronomy, School of Physics and Astronomy, Key Laboratory of Astroparticle Physics of Yunnan Province, Yunnan University, Kunming 650091, China; {\it lizhang@ynu.edu.cn}\\
\and
Southwest United Graduate School, Kunming 650092, China\\
\vs\no
{\small Received 2026 month day; accepted 2026 month day}}

\abstract{We present a reanalysis of 17 years of \textit{Fermi} Large Area Telescope (LAT) observations of the Crab pulsar obtained between 2008 August and 2025 August. Using monthly Jodrell Bank radio ephemerides, we assigned pulse phases to the LAT events and aligned the phase zero across the full dataset. From this phase-aligned dataset, we derived pulse profiles over 100 MeV$-$300 GeV, which remain clearly detectable in the 10$-$20 and 20$-$30 GeV bands with H-test significances of 32.36$\sigma$ and 11.59$\sigma$, respectively, but are not significantly detected in the 30$-$300 GeV band. Phase-resolved likelihood analysis was performed over 100 MeV$-$30 GeV using 14 phase bins with comparable pulsed statistics. The fixed-window fractional fluxes show that the contribution of Peak 1 (P1) decreases steadily with energy, while those of Peak 2 (P2) and the Bridge increase, with P2 exceeding P1 above 10 GeV. Finally, the same phase-assignment framework also enables an off-pulse analysis from 100 MeV to 1 TeV, confirming the synchrotron and IC components that dominate the emission in the selected off-pulse interval.
}

\authorrunning{L. Zhou et al.}
\titlerunning{Gamma-Ray Emission from the Crab Pulsar}
\maketitle

\section{Introduction}
The Crab system, consisting of the Crab pulsar and its surrounding nebula, remains a benchmark source for studying broadband radiation across the electromagnetic spectrum. Located at a distance of about 2 kpc, this system provides a unique opportunity to investigate both pulsed and nebular emission. Pulsed emission from the Crab pulsar has been detected across nearly the entire spectrum, from radio waves to very-high-energy gamma rays, with secure pulsed emission detected above 100 GeV \citep{the_veritas_collaboration_detection_2011, aleksic_phase-resolved_2012, ansoldi_teraelectronvolt_2016}. In addition, the alignment of peak phases in the $\gamma$-ray and radio light curves was first established in the Crab pulsar \citep{1999ApJ...516..297T, abdo_fermi_2010}. Therefore, the Crab system has played a crucial role in advancing our understanding of high-energy emission of the pulsars.

For the \textit{Fermi}-LAT analysis of the Crab pulsar, strong timing noise and glitches make it difficult to construct a single phase-coherent ephemeris covering the full time span \citep{kerr_timing_2015,pshirkov_gamma-ray_2020}. We address this problem by assigning phases with the public monthly Crab radio ephemerides provided by Jodrell Bank Observatory\citep{1993MNRAS.265.1003L} and aligning the phase zero across segments. This procedure yields a phase-aligned \textit{Fermi}-LAT data set covering 17 years and enables the reanalysis presented here. Earlier \textit{Fermi}-LAT studies established the GeV pulse morphology, phase alignment, and broadband spectral properties of the Crab pulsar as well as GeV emission from the Crab nebula \citep{abdo_fermi_2010}. Later work examined how the pulsed and bridge emission change with energy \citep{yeung_inferring_2020}. More recently, a joint analysis combining \textit{Fermi}-LAT and CTAO LST-1 data provided a detailed phase-resolved study of the Crab pulsar and further constrained the peak and bridge emission at very high energies \citep{2024A&A...690A.167A}.

In this paper, we use the 17-year phase-aligned \textit{Fermi}-LAT data set to revisit the Crab pulsar through pulse-profile analysis over 100 MeV$-$300 GeV, equal-pulsed-statistics phase-resolved spectral analysis over 100 MeV$-$30 GeV, and off-pulse nebular analysis over 100 MeV$-$1 TeV. We also describe in detail the data processing and analysis procedure adopted in this work. Section~\ref{sec:data_phase_assignment} describes the LAT data selection, phase assignment and alignment, and pulse-profile construction, Section~\ref{sec:phase_resolved} presents the energy-resolved pulse profiles, phase-resolved spectra, and the energy dependence of the fixed-window fractional fluxes, and Section~\ref{sec:offpulse_pwn} presents the off-pulse analysis. Section~\ref{sec:summary_discussion} summarizes and discusses the main results.

\section{The Phase-Aligned Fermi-LAT Dataset of the Crab Pulsar}
\label{sec:data_phase_assignment}
\subsection{LAT Data Selection}
\label{subsec:lat_data_selection}
We analyzed \textit{Fermi}-LAT data spanning 2008 August 4 17:00:58.140 to 2025 August 5 23:58:23.101 (UTC), corresponding to MET 239562059.14$-$776131108.10.
The data were downloaded from the Fermi Science Support Center (FSSC)\footnote{\url{https://fermi.gsfc.nasa.gov/ssc/data/access/}} and processed according to LAT analysis recommendations\footnote{\url{https://fermi.gsfc.nasa.gov/ssc/data/analysis/documentation/Cicerone/}}.

We selected Pass~8 SOURCE-class events with \texttt{evclass=128} and \texttt{evtype=3} (front+back conversion types) in the 100 MeV$-$300 GeV energy range. The events were extracted from a 3$^{\circ}$ region centered on \texttt{4FGL J0534.5+2200} in the 4FGL-DR4 catalog \citep{ballet_fermi_2024} at $(\alpha,\delta)=({05^{\mathrm h}34^{\mathrm m}32\fs0}, {+22^\circ00'52\farcs1})$ (J2000). To suppress Earth-limb contamination and ensure good-quality observing intervals, we applied \texttt{zmax=90$^{\circ}$}, \texttt{DATA\_QUAL>0}, and \texttt{LAT\_CONFIG==1}.

\subsection{Phase Assignment and Alignment}
\label{subsec:phase_assignment_alignment}
Phase assignment over 17 years of LAT data is particularly challenging for the Crab pulsar because strong timing noise and glitches are present over such a long time span, making it difficult to obtain a single phase-coherent ephemeris that reliably spans the full dataset \citep{pshirkov_gamma-ray_2020}. For phase assignment, we adopted the public monthly Crab radio ephemerides provided by Jodrell Bank Observatory \citep{1993MNRAS.265.1003L}\footnote{\url{https://www.jb.man.ac.uk/pulsar/crab.html}} and converted each validity interval into a timing solution containing the parameters \texttt{START}, \texttt{FINISH}, \texttt{PEPOCH}, \texttt{F0}, \texttt{F1}, and \texttt{F2}, following established LAT pulsar timing conventions \citep{kerr_timing_2015}.

Using the MJD and \texttt{t\_JPL} values listed in the public Jodrell Bank Crab timing results\footnote{\url{https://www.jb.man.ac.uk/pulsar/crab/crab2.txt}}, we determined the arrival time of the radio main pulse for each monthly segment and used it as the reference epoch for phase assignment. Here, \texttt{t\_JPL} gives the arrival-time offset (in seconds) of the radio main pulse relative to the listed MJD epoch. In this way, all LAT events were aligned to a common phase-zero definition across the full data span.

For each monthly validity interval, we ran \texttt{gtpphase}\footnote{\url{https://fermi.gsfc.nasa.gov/ssc/data/analysis/scitools/pulse_phase_tutorial.html}}\footnote{\url{https://fermi.gsfc.nasa.gov/ssc/data/analysis/scitools/references.html}} on FT1 events with the corresponding FT2 spacecraft file and monthly timing solution, using \texttt{FREQ} mode with \texttt{timesys=TDB}, \texttt{tcorrect=BARY}, and \texttt{solareph=JPL~DE200}. This procedure assigns \texttt{PULSE\_PHASE} to each event and yields the phase-aligned FT1 files. We then combined the monthly outputs into a single phase-aligned dataset for the subsequent analyses.

\subsection{Pulse Profiles and Phase-Interval Definitions}
\label{subsec:pulse_profiles}
\begin{figure*}
	\centering
	\includegraphics[width=0.95\textwidth]{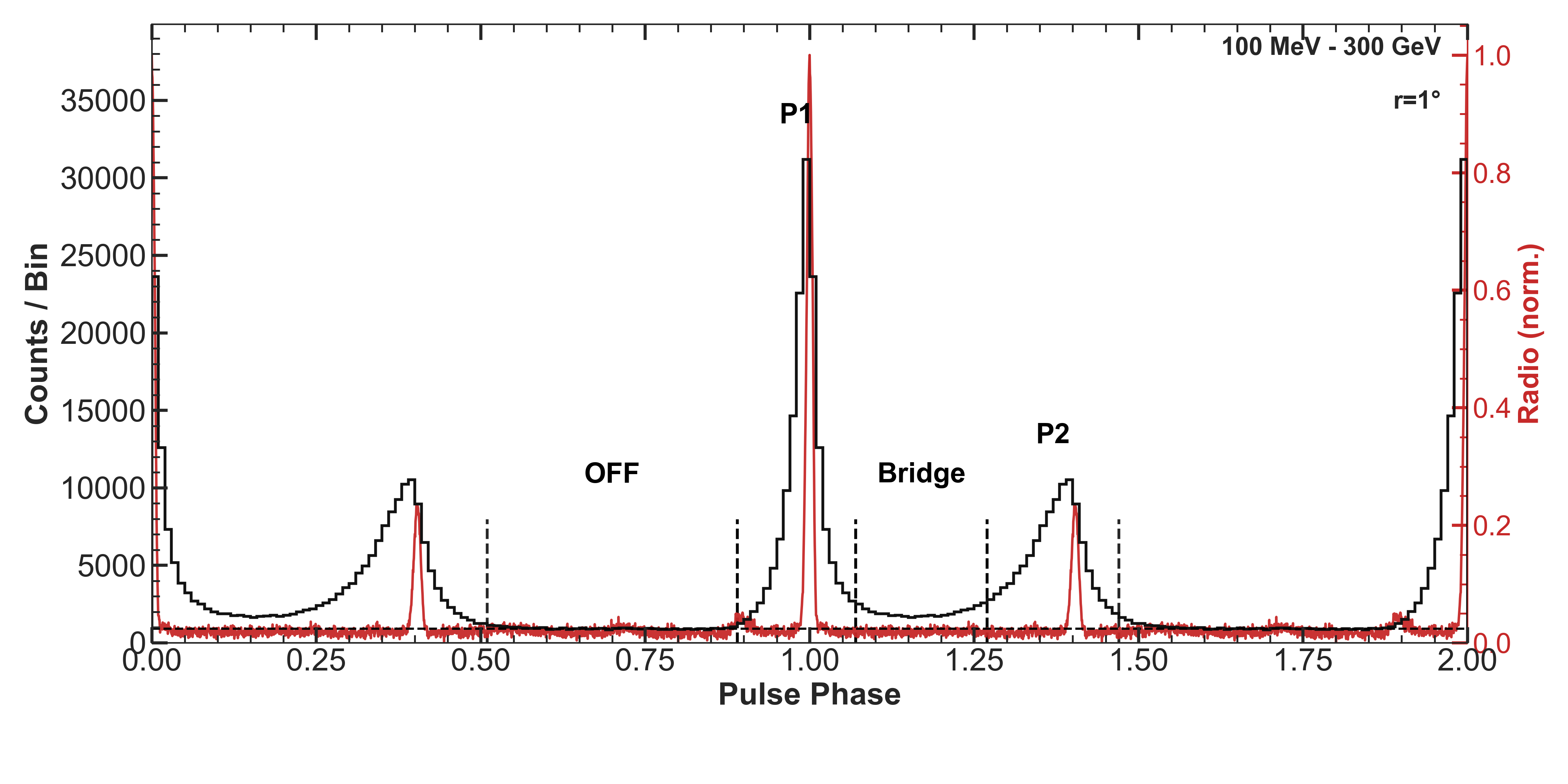}
	\caption{Integrated pulse profile of the Crab pulsar constructed from the phase-aligned FT1 events in the 100 MeV$-$300 GeV energy range within a $1^\circ$ radius. The profile is shown over two phase cycles (0$-$2) for visual continuity and is histogrammed into 100 phase bins. The black step histogram gives the counts per phase bin, and the black dashed horizontal line represents the estimated background level. The red curve shows the normalized radio profile from the Nancay telescope at 1.4 GHz, referenced to the right-hand axis, and illustrates the alignment of the radio main pulse with the adopted phase-zero definition. Black dashed vertical lines show the boundaries of the phase intervals labeled P1, Bridge, P2, and OFF.}
	\label{fig:phase_windows}
\end{figure*}

From the merged phase-aligned FT1 dataset, we constructed an integrated gamma-ray pulse profile in the 100 MeV$-$300 GeV energy range within a $1^\circ$ radius and defined the phase intervals used throughout the analysis, including fixed Peak 1 (P1), Peak 2 (P2), Bridge, and off-pulse intervals. The integrated profile is shown in Figure~\ref{fig:phase_windows} and was histogrammed into 100 phase bins. Figure~\ref{fig:phase_windows} also includes a normalized radio profile\footnote{The radio profile is available from the LAT pulsar ephemerides page: \url{https://fermi.gsfc.nasa.gov/ssc/data/access/lat/ephems/}.} in red, referenced to the right-hand axis, to illustrate the phase alignment between the radio main pulse and the gamma-ray peaks. For the two main pulses, we adopted fixed phase windows broadly following \citet{abdo_fermi_2010}, with P1 in the 0.89$-$1.07 phase range and P2 in the 0.27$-$0.47 phase range. In the present work, the leading edge of P1 was shifted from 0.87 to 0.89 so that the fixed peak window does not overlap with the off-pulse interval defined below. The off-pulse interval was determined independently from the data using a quasi-baseline (QB) procedure motivated by \citet{meyer_characterizing_2019}: we estimated the baseline level $B_0$ and dispersion $S_0$ from the optimally symmetric part of the phase-histogram count distribution, selected off-peak bins with counts $\leq B_0+6.5\,S_0$, and retained a single contiguous component on the phase circle as the final off-pulse interval. For the integrated pulse profile in the 100 MeV$-$300 GeV range, this procedure yielded an off-pulse interval of 0.51$-$0.89 in phase, consistent with the off-pulse phase interval adopted in recent broadband Crab pulsar analysis \citep{aharonian_spectrum_2024}. We define the ON region as the phase range outside the OFF region and use these ON/OFF regions in the following analysis. The Bridge region is taken as the main inter-peak segment 0.07$-$0.27, as shown in Figure~\ref{fig:phase_windows}. These fixed P1, P2, and Bridge windows are also used in the following fixed-window SED and energy-dependent fractional-flux analysis.

\subsubsection{Energy-Resolved Profiles}
\label{subsubsec:energy_resolved_profiles}
\begin{figure*}
	\centering
	\includegraphics[width=0.95\textwidth,height=0.68\textheight,keepaspectratio]{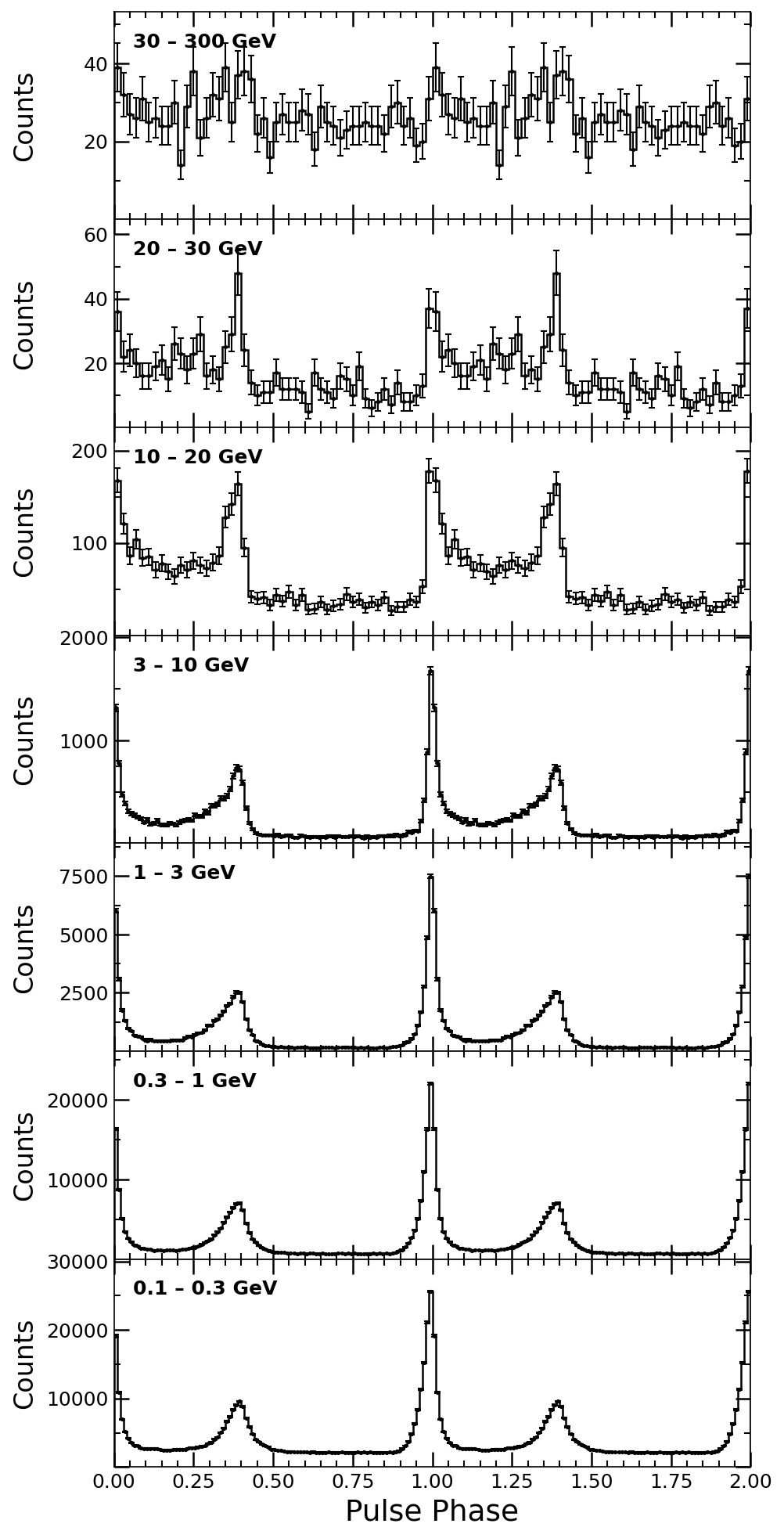}
	\caption{Energy-resolved Crab pulsar pulse profiles constructed from the merged phase-aligned FT1 dataset using the energy-dependent angular cut described in Section~\ref{subsubsec:energy_resolved_profiles}. The profiles are shown over two phase cycles (0$-$2) and are ordered from top to bottom as 30$-$300, 20$-$30, 10$-$20, 3$-$10, 1$-$3, 0.3$-$1, and 0.1$-$0.3 GeV. No event weights were applied in this figure, so the y-axis shows raw counts ($N$) and the error bars correspond to Poisson $\sqrt{N}$ uncertainties per phase bin. Profiles below 10~GeV use 100 phase bins, while the 10$-$20, 20$-$30, and 30$-$300~GeV panels use 50 phase bins to reduce low-count fluctuations.}
	\label{fig:profiles_main}
\end{figure*}

For event selection, we applied the energy-dependent angular-cut approach of \citet[Section~4.1]{abdo_fermi_2010}, with
\begin{equation}
\theta_{\mathrm{cut}}(E) = \max\left(6.68 - 1.76\log_{10}\frac{E}{\mathrm{MeV}},\ 1.3\right)\ \mathrm{deg}.
\end{equation}
and then constructed the energy-resolved pulse profiles from the selected events.
For $E\geq10$~GeV, we adopted 50 phase bins to stabilize low-count fluctuations. For the profiles shown in Figure~\ref{fig:profiles_main}, no event weights were applied ($w_i=1$), so the error bars were computed as Poisson $\sqrt{N}$ uncertainties in each phase bin.

Figure~\ref{fig:profiles_main} shows the energy-resolved pulse profiles obtained with this energy-dependent angular cut. The pulse profiles remain clearly visible in the 10$-$20~GeV and 20$-$30~GeV bands. Table~\ref{tab:high_energy_stats} summarizes the event counts and pulse-profile statistics for all energy bands. For each band, it lists the selected event count $N_{\rm evt}$, the unweighted H-test statistic\citep{2010A&A...517L...9D}, and its Gaussian-equivalent significance computed by PINT \citep{2021ApJ...911...45L}. In particular, the 20$-$30~GeV band retains an H-test significance of 11.59$\sigma$ (Table~\ref{tab:high_energy_stats}), benefiting from the nearly 17-year \textit{Fermi}-LAT exposure used in this work and the resulting increase in photon statistics.

\begin{table}[t]
\centering
\footnotesize
\caption{Pulse-profile statistics for the energy bands shown in Figure~\ref{fig:profiles_main}, computed from the same unweighted, phase-assigned FT1 sample with the energy-dependent angular cut. Column definitions and calculation details are given in Section~\ref{subsubsec:energy_resolved_profiles}.\label{tab:high_energy_stats}}
\begin{tabular*}{\columnwidth}{@{\extracolsep{\fill}}cccc@{}}
\hline
Band (GeV) & $N_{\rm evt}$ & H-test & Significance ($\sigma$) \\
\hline
0.1$-$0.3   & 415890 & $3.77{\times}10^{5}$ & 613.45 \\
0.3$-$1     & 247642 & $4.76{\times}10^{5}$ & 690.00 \\
1$-$3       & 79979  & $1.68{\times}10^{5}$ & 409.22 \\
3$-$10      & 22298  & $2.89{\times}10^{4}$ & 169.48 \\
10$-$20     & 3204   & 1143.88              & 32.36 \\
20$-$30     & 839    & 170.18               & 11.59 \\
30$-$300    & 1333   & 6.87                 & 1.85 \\
\hline
\end{tabular*}
\end{table}

\section{Phase-Resolved Spectral Analysis and Pulse Statistics}
\label{sec:phase_resolved}
\subsection{Equal-Pulsed-Statistics Phase Binning}
\label{subsec:eq_pulsed_binning}
Using the 17-year \textit{Fermi}-LAT phase-aligned FT1 dataset constructed in Section~\ref{sec:data_phase_assignment}, we defined phase intervals with comparable pulsed statistics for the phase-resolved SED analysis. The phase-bin boundaries were built from the event sample selected with the energy-dependent angular cut described in Section~\ref{subsubsec:energy_resolved_profiles}, after estimating the pulsed-count distribution in the ON region and partitioning its cumulative distribution into 14 bins with equal pulsed-count targets. For each phase element \(i\) with width \(\Delta\phi_i\), \(N_i\) denotes the observed event counts in that element, and the estimated pulsed counts are
\begin{equation}
N_{\mathrm{pulsed},i}=\max\left(N_i-\rho_{\mathrm{bkg}}\Delta\phi_i,0\right).
\end{equation}
Here, \(\rho_{\mathrm{bkg}}\) is estimated from the OFF region as \(\rho_{\mathrm{bkg}}=N_{\mathrm{OFF}}/\Delta\phi_{\mathrm{OFF}}\), where \(N_{\mathrm{OFF}}\) is the number of selected events in the OFF region and \(\Delta\phi_{\mathrm{OFF}}\) is the OFF phase width. The term \(\rho_{\mathrm{bkg}}\Delta\phi_i\) gives the expected background counts in phase element \(i\). We then accumulated \(N_{\mathrm{pulsed},i}\) across the ON region and set phase boundaries at equal cumulative pulsed counts to define the 14 analysis bins. Similar equal-pulsed-statistics phase-binning ideas have been used in previous LAT pulsar works \citep{abdo_fermi_2010,smith_third_2023,lange_fermiphased_2025}. These bins were used as fixed inputs in the subsequent likelihood fits.

\subsection{Likelihood Models and SED Extraction}
\label{subsec:likelihood_sed_extraction}
Each phase bin was fitted independently with a binned maximum-likelihood analysis. The fits used the same data time span and baseline event-selection criteria as Section~\ref{subsec:lat_data_selection} (\texttt{evclass=128}, \texttt{evtype=3}, \(z_{\max}=90^\circ\), and \texttt{(DATA\_QUAL>0)\&\&(LAT\_CONFIG==1)}), with the source model initialized from the 4FGL-DR4 catalog (\texttt{gll\_psc\_v33.fit}). Phase selections were applied with direct phase-window cuts using the phase intervals defined in Section~\ref{subsec:eq_pulsed_binning}, and the exposure normalization in each fit was scaled by the corresponding phase width. The spectral fits used the ROI event selection in the 100 MeV$-$30 GeV energy range. Before running the phase-resolved fits, we first fitted the OFF region and then used the resulting best-fit background parameters for 4FGL J0534.5+2201s, 4FGL J0534.5+2201i, and the Galactic and isotropic diffuse components as fixed inputs in each phase-bin fit to reduce background influence.

In each bin, we compared a pure power-law (PL) model and a super-exponential cutoff power-law (PLEC4) model\footnote{\url{https://fermi.gsfc.nasa.gov/ssc/data/analysis/scitools/source_models.html}}. We followed the PLEC4 model (\texttt{PLSuperExpCutoff4}; see \citet{abdollahi_incremental_2022} for details, and see also Eq.~(15) of \citet{smith_third_2023}). The PLEC4 cutoff-shape parameter \(b\) was left free and constrained to \(1/3\leq b\leq 4/3\), following the recommendation of \citet{smith_third_2023}.

Spectral energy distribution (SED) points were extracted with bin-by-bin likelihood. For each of the 14 equal-pulsed phase intervals, we used six logarithmic energy bins. For the fixed-window P1/P2/Bridge/whole-pulse spectra shown in Figure~\ref{fig:whole_pulse_sed}, we used eight logarithmic energy bins over the same 100 MeV$-$30 GeV range. We adopted a TS value of \(<4\) as the upper-limit criterion. For model comparison, we retained only converged fits with acceptable quality and used the Akaike information criterion \citep[AIC;][]{1974ITAC...19..716A} as the primary metric:
\begin{equation}
\mathrm{AIC}=2k-2\ln\hat{\mathcal L},
\end{equation}
where \(k\) is the number of free parameters and \(\hat{\mathcal L}\) is the maximized likelihood value. For compact reporting, we define
\begin{equation}
\Delta\mathrm{AIC}\equiv \mathrm{AIC}_{\rm PL}-\mathrm{AIC}_{\rm PLEC4},
\end{equation}
so positive values favor PLEC4. The model-comparison results are summarized in Table~\ref{tab:phase_params_placeholder}.

\subsection{Phase-Resolved SEDs}
\label{subsec:phase_sed_results}
For the phase-resolved likelihood and SED fits, we adopted a single response and diffuse-background configuration: the Pass 8 SOURCE-class instrument response functions (IRFs; \texttt{P8R3\_SOURCE\_V3}), the Galactic diffuse model \texttt{gll\_iem\_v07.fits}, and the isotropic diffuse model \texttt{iso\_P8R3\_SOURCE\_V3\_v1.txt}. Energy dispersion was enabled for source components and disabled for the diffuse components (\texttt{galdiff} and \texttt{isodiff}). This configuration follows current Fermi-LAT analysis recommendations \citep{wood_fermipy_nodate,abdollahi_fermi_2020,ballet_fermi_2024}.

IRF systematic uncertainties were evaluated with bracketing-response reruns, following the official FSSC Aeff-systematics guidance\footnote{\url{https://fermi.gsfc.nasa.gov/ssc/data/analysis/scitools/Aeff_Systematics.html}}. For each phase bin, we constructed minimum and maximum effective-area scaling curves, sampled with 15 logarithmic points per decade across the IRF-bracketing energy domain used in this analysis. We then reran the optimization and likelihood fit for each bracketing case and compared the resulting SED with the nominal fit. This procedure provided upper and lower systematic shifts, \(\sigma_{\mathrm{sys},+}\) and \(\sigma_{\mathrm{sys},-}\), at each energy bin. To report one systematic term per bin, we adopted
\begin{equation}
\sigma_{\mathrm{sys}}=\max\!\left(\sigma_{\mathrm{sys},+},\sigma_{\mathrm{sys},-}\right),
\end{equation}
We then combined this systematic term with the statistical uncertainty:
\begin{equation}
\sigma_{\mathrm{tot}}=\sqrt{\sigma_{\mathrm{stat}}^2+\sigma_{\mathrm{sys}}^2}.
\end{equation}

\begin{figure}
	\centering
	\includegraphics[width=\columnwidth]{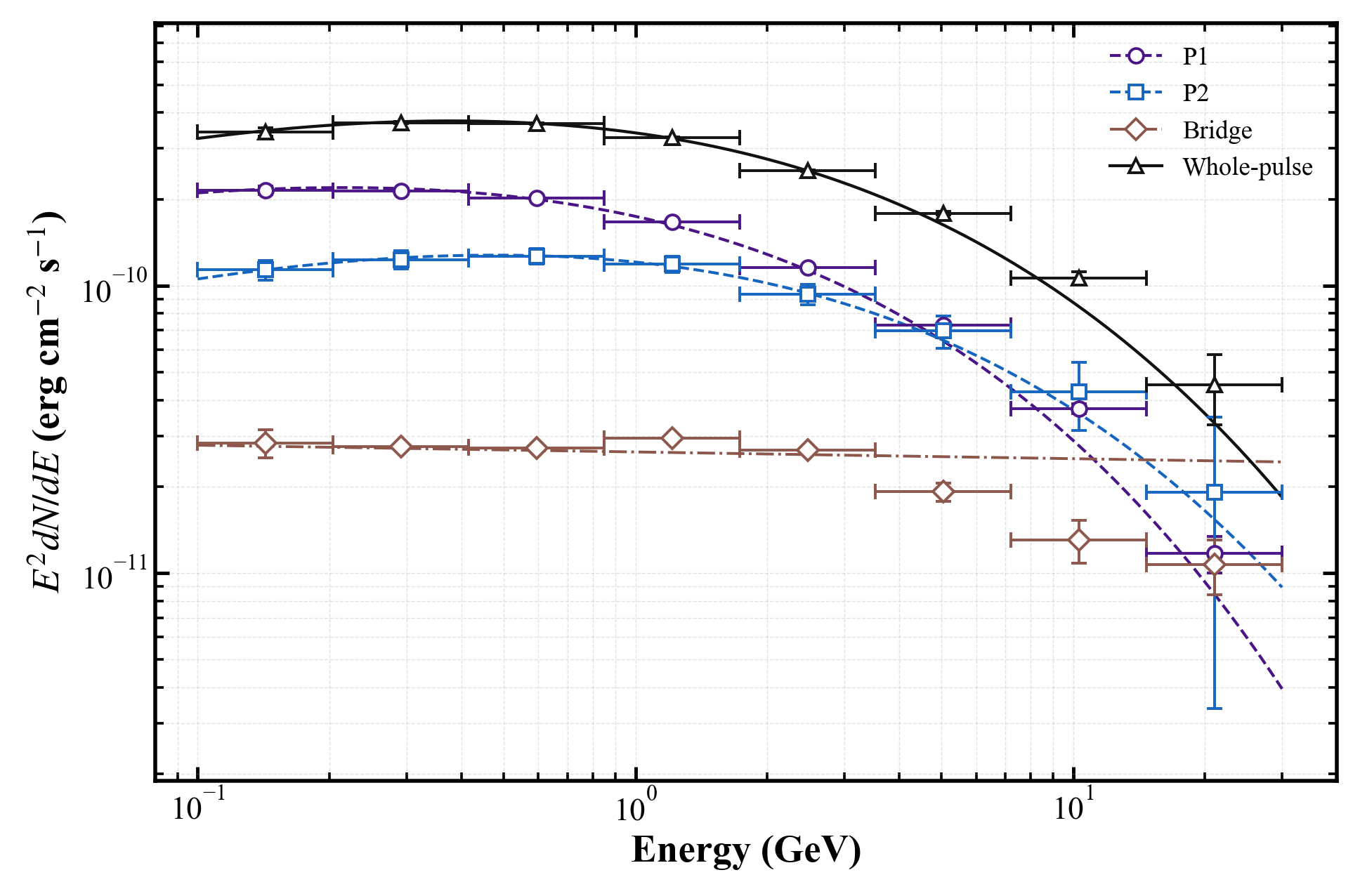}
			\caption{Phase-integrated SEDs for the fixed P1, P2, and Bridge windows, together with the full phase interval, in the 100 MeV$-$30 GeV fitting range. Here the Bridge window is the 0.07$-$0.27 interval. The fixed-window SED points are shown with eight logarithmic energy bins over 100 MeV$-$30 GeV, and no additional display-level merging was applied to the Bridge points. In this figure, the displayed model curves are PLEC4 for P1, P2, and the whole-pulse spectrum, and PL for the Bridge spectrum. The plotted uncertainties are total uncertainties after including the estimated systematics.}
	\label{fig:whole_pulse_sed}
\end{figure}

With this uncertainty treatment, Figure~\ref{fig:phase_sed_summary_31}, together with its continued panel, shows the phase-resolved spectral energy distributions for the 14 equal-pulsed phase bins.
Figure~\ref{fig:whole_pulse_sed}
shows the phase-integrated contribution SEDs for the fixed P1, P2, and Bridge windows and the full phase interval defined in Section~\ref{subsec:pulse_profiles}. In this figure, the Bridge window corresponds to the 0.07$-$0.27 interval. The fixed-window SED points shown here were extracted with eight logarithmic energy bins over 100 MeV$-$30 GeV, and no additional display-level merging was applied to the Bridge points. For visual consistency in this figure, the displayed model curves are PLEC4 for P1, P2, and the whole-pulse spectrum, and PL for the Bridge spectrum. Together, these spectra illustrate the relative contributions of the different phase intervals to the whole-pulse emission.
Table~\ref{tab:phase_params_placeholder}
summarizes the 14 equal-pulsed phase intervals, their per-unit-phase spectral properties, and the AIC-based model preference.
Using the \(\Delta\mathrm{AIC}\) definition introduced in Section~\ref{subsec:likelihood_sed_extraction}, PLEC4 was preferred over PL for the whole-pulse spectrum.

From the whole-pulse fit, we derived the following quantities:
\begin{equation}
\Gamma_{100}=\gamma_0-\frac{d}{b}+\frac{d}{b}\left(\frac{100\ \mathrm{MeV}}{E_0}\right)^b,
\end{equation}
\begin{equation}
E_p=E_0\left[1+\frac{b(2-\gamma_0)}{d}\right]^{1/b},
\end{equation}
and
\begin{equation}
d_p=d+b(2-\gamma_0).
\end{equation}
where \(E_0\) is the scale energy and \(\Gamma_{100}\) is the spectral slope at 100 MeV. In the Third \textit{Fermi}-LAT Pulsar Catalog (3PC)\citep{smith_third_2023}, the reported values for the Crab pulsar are: the 0.1$-$100 GeV energy flux \(G_{100}=(150.0\pm1.3)\times10^{-11}\ \mathrm{erg\,cm^{-2}\,s^{-1}}\), \(\Gamma_{100}=2.00\pm0.03\), \(E_p=0.04\pm0.03\ \mathrm{GeV}\), \(d_p=0.23\pm0.01\), and \(b=0.61\pm0.08\). In this work, for the 100 MeV$-$30 GeV fit, we obtained the energy flux \(G=(141.3\pm0.4)\times10^{-11}\ \mathrm{erg\,cm^{-2}\,s^{-1}}\), \(\Gamma_{100}=1.92\pm0.01\), \(E_p=0.30\pm0.02\ \mathrm{GeV}\), \(d_p=0.10\pm0.01\), and \(b=0.56\pm0.03\). The uncertainties on \(\Gamma_{100}\), \(E_p\), and \(d_p\) were estimated with first-order error propagation from the fitted statistical errors on \(\gamma_0\), \(d\), and \(b\).

\begin{figure*}
	\centering
	\includegraphics[width=0.42\textwidth]{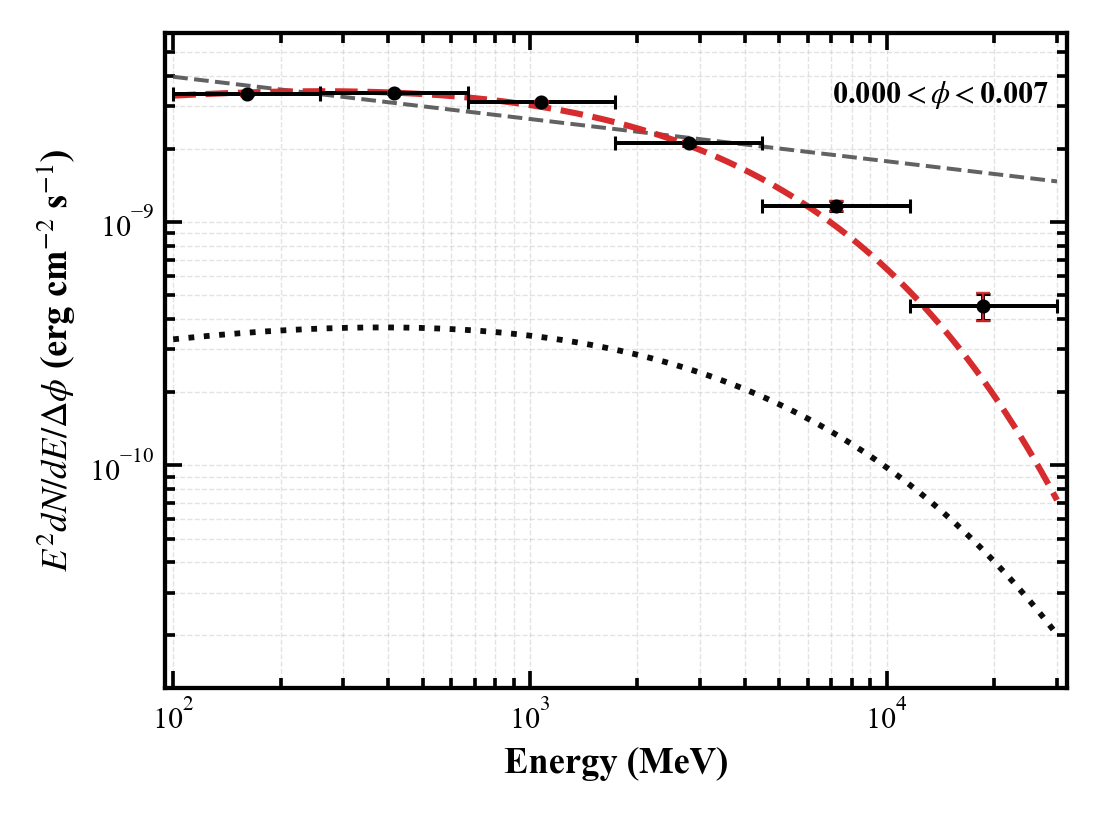}\hspace{0.012\textwidth}
	\includegraphics[width=0.42\textwidth]{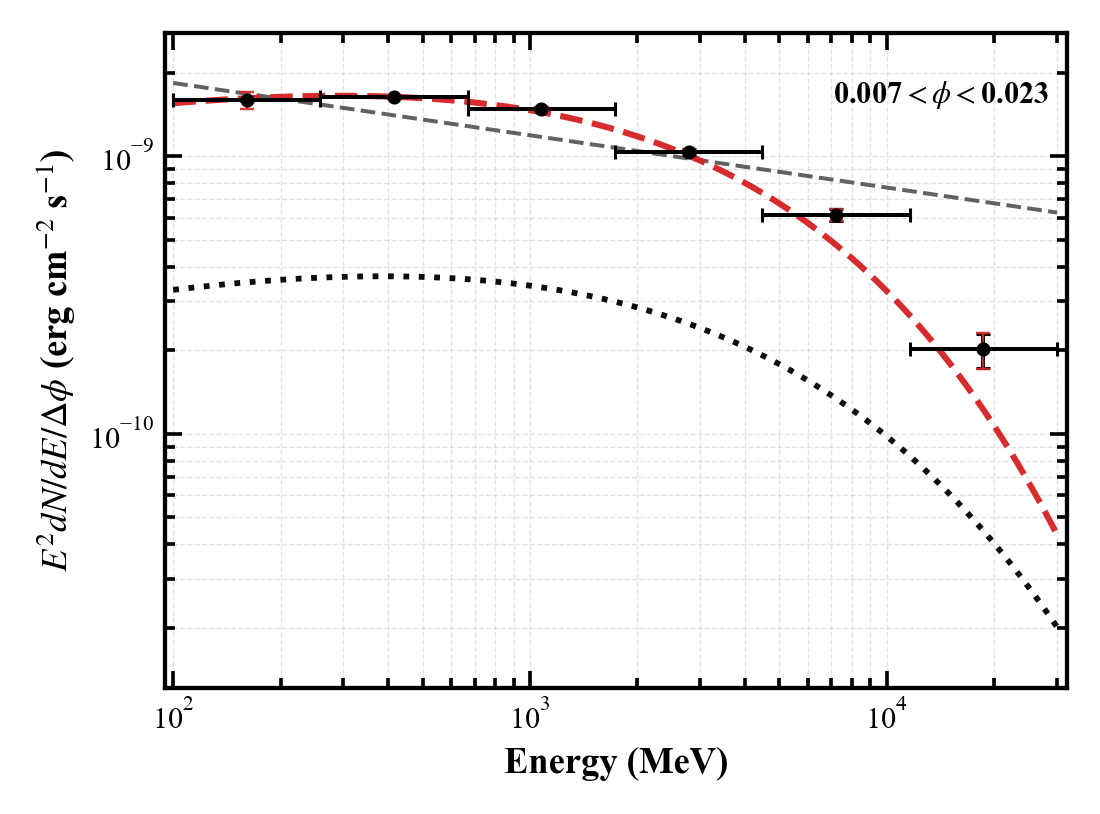}\\[-4pt]
	\includegraphics[width=0.42\textwidth]{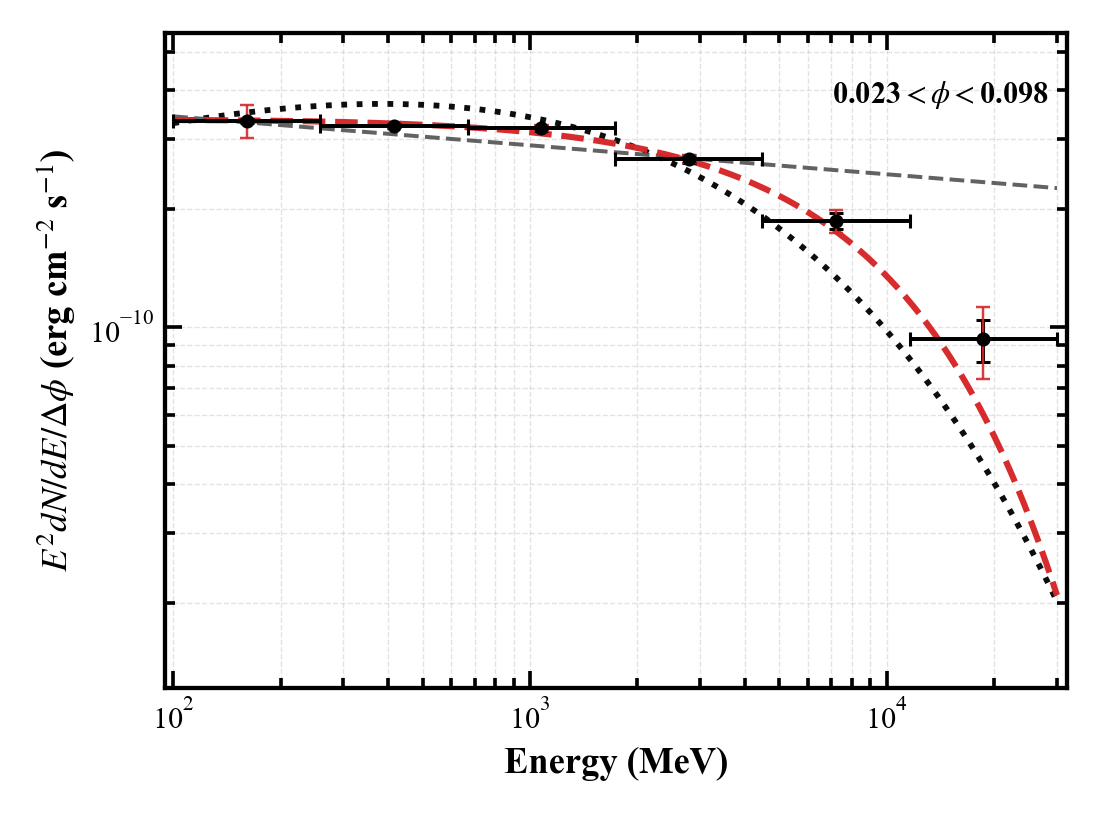}\hspace{0.012\textwidth}
	\includegraphics[width=0.42\textwidth]{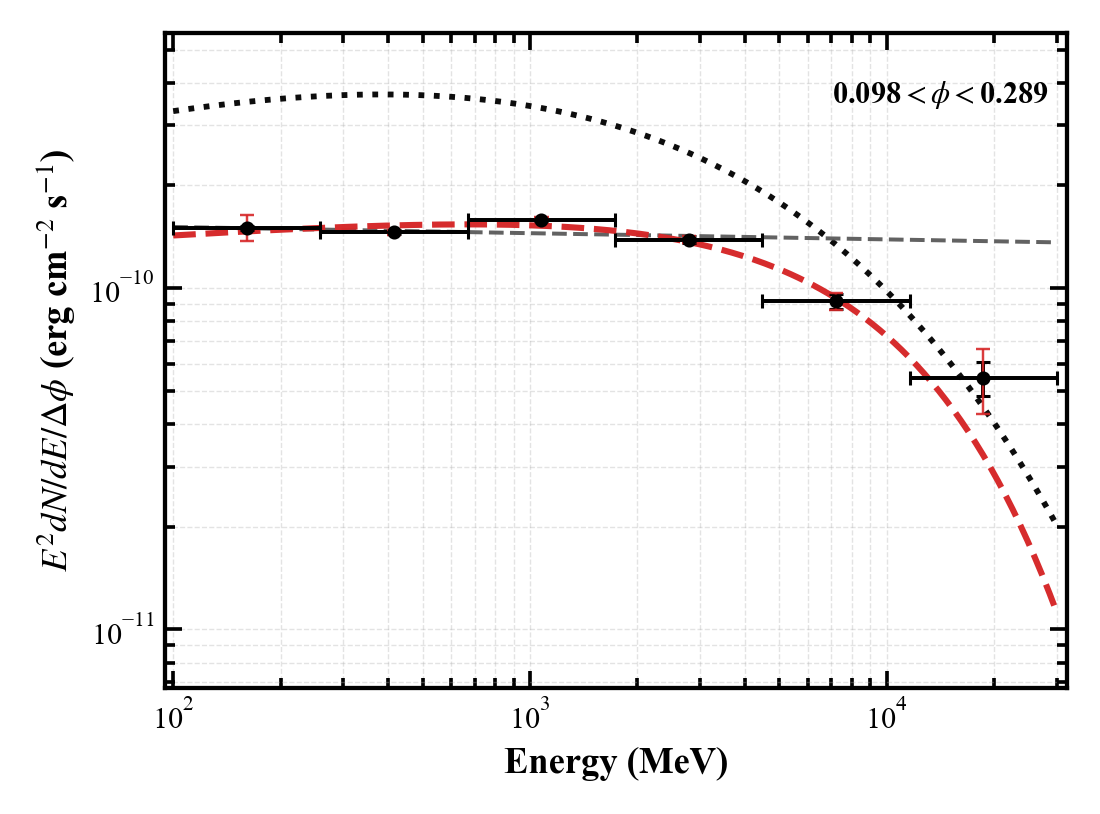}\\[-4pt]
	\includegraphics[width=0.42\textwidth]{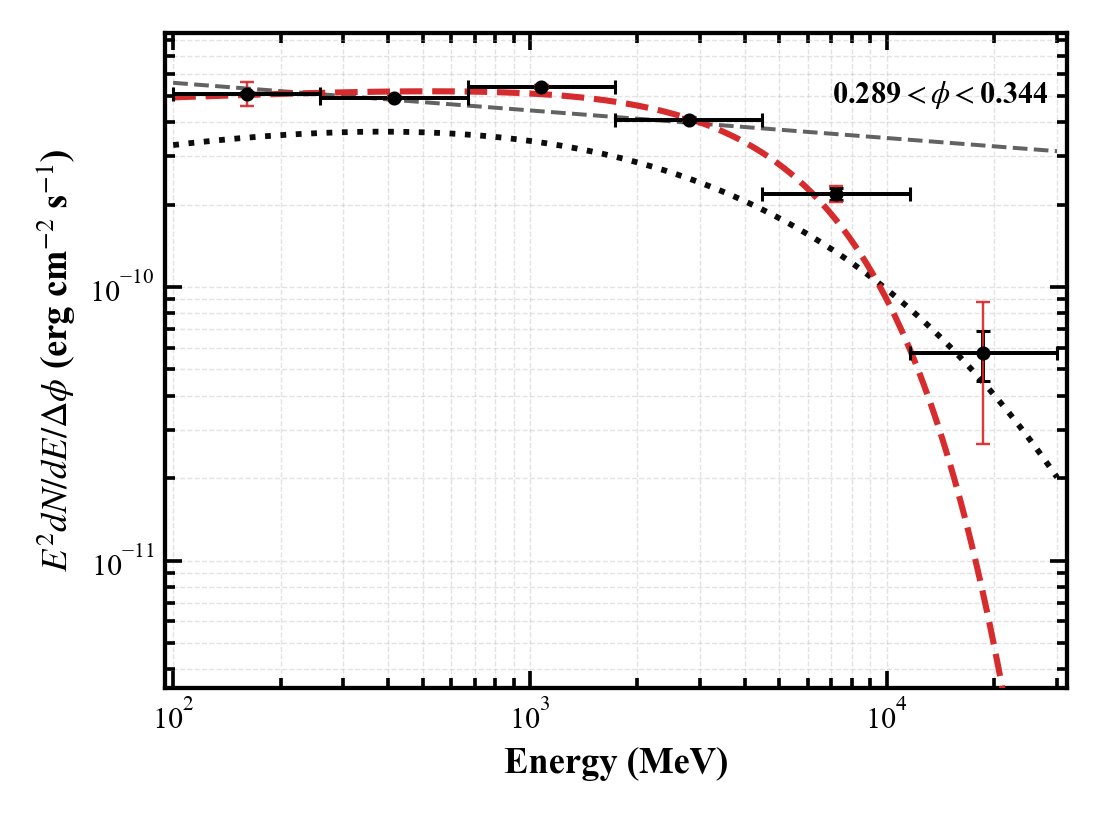}\hspace{0.012\textwidth}
	\includegraphics[width=0.42\textwidth]{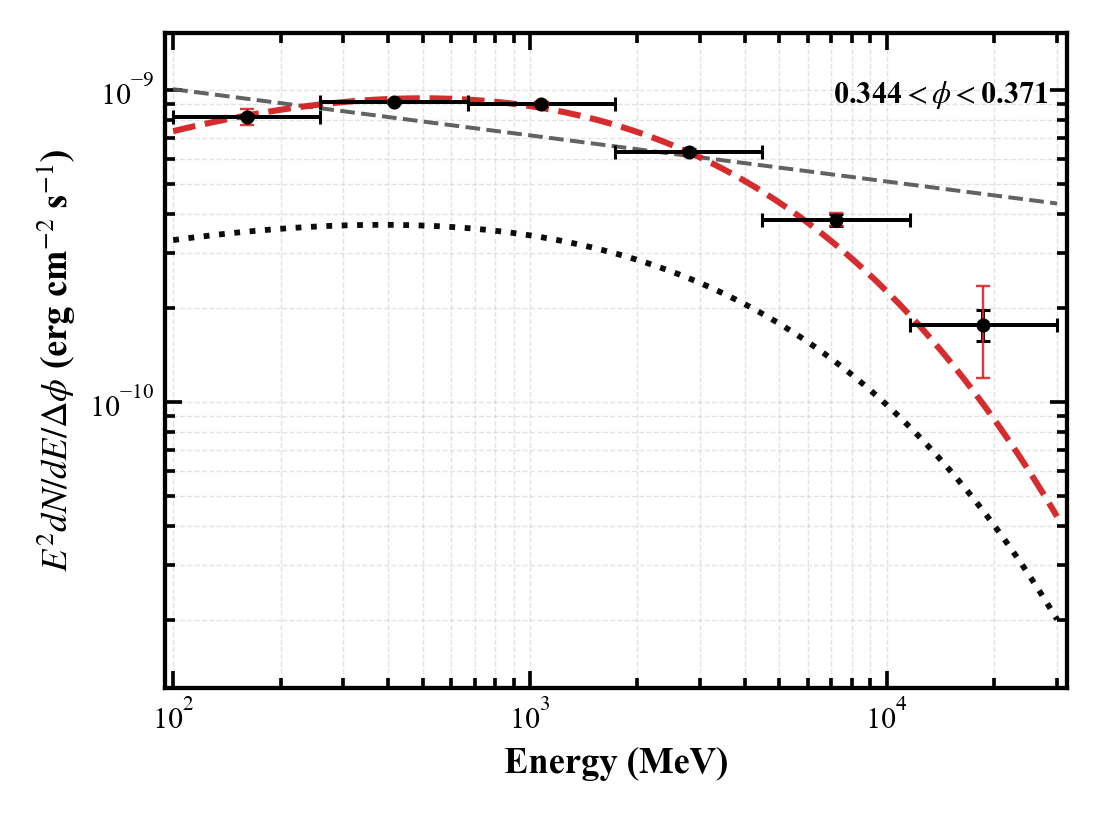}\\[-4pt]
	\includegraphics[width=0.42\textwidth]{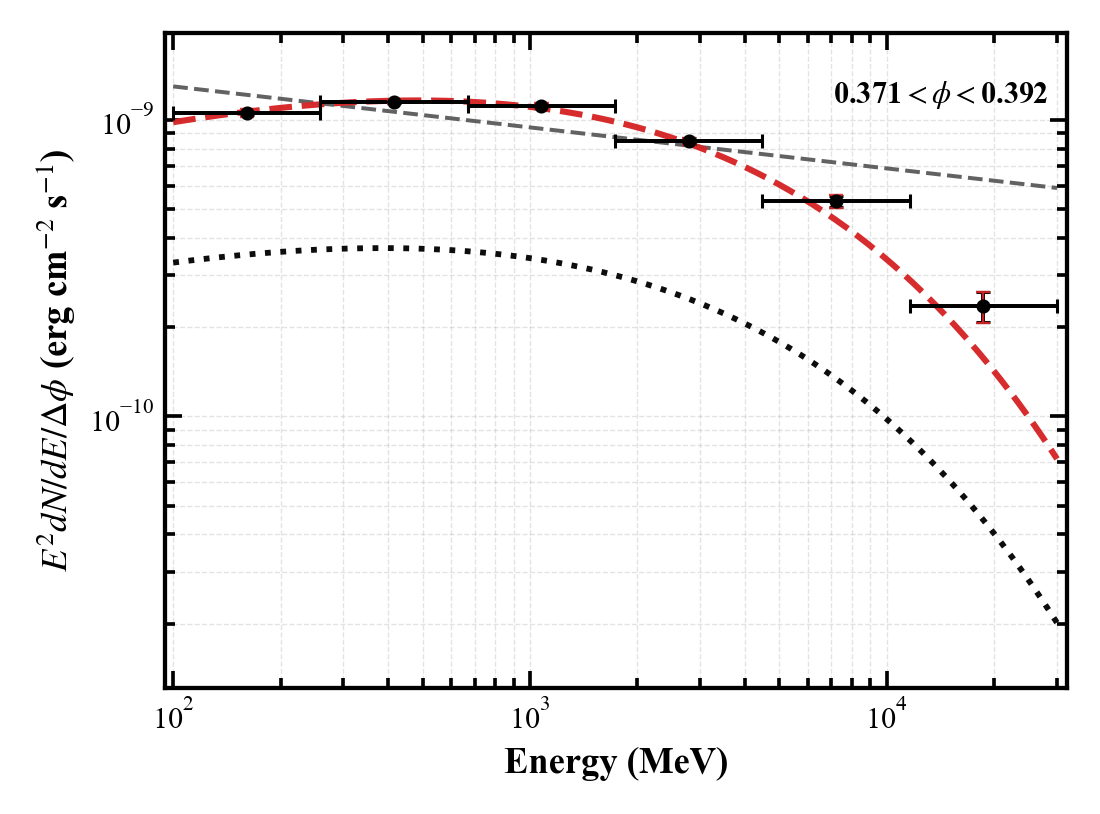}\hspace{0.012\textwidth}
	\includegraphics[width=0.42\textwidth]{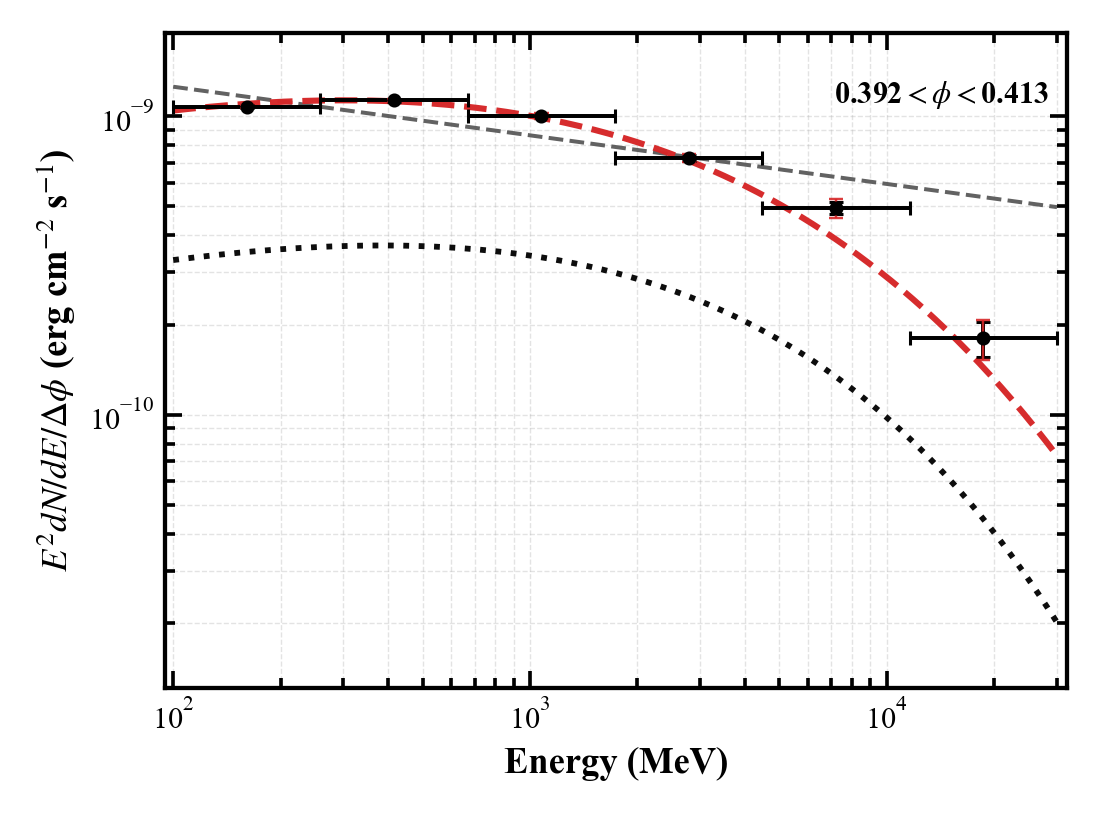}
	\vspace{-1pt}
			\caption{Phase-resolved spectral energy distributions of the Crab pulsar for phase bins 00$-$07 in the 100 MeV$-$30 GeV fitting range (two panels per row, four rows). The plotted SED quantity is normalized by the corresponding phase width, \(E^2 dN/dE/\Delta\phi\), consistent with the flux convention used in Table~\ref{tab:phase_params_placeholder}. The labels indicate the phase intervals, and the corresponding spectral results are presented in Table~\ref{tab:phase_params_placeholder}. Each panel compares the \texttt{PowerLaw} and \texttt{PLSuperExpCutoff4} fits; black points show statistical uncertainties, and red markers show total uncertainties after including the estimated systematics. The dotted black curve shows the whole-pulse PLEC4 spectral line for comparison.}
	\label{fig:phase_sed_summary_31}
\end{figure*}

\begin{figure*}
	\centering
	\includegraphics[width=0.42\textwidth]{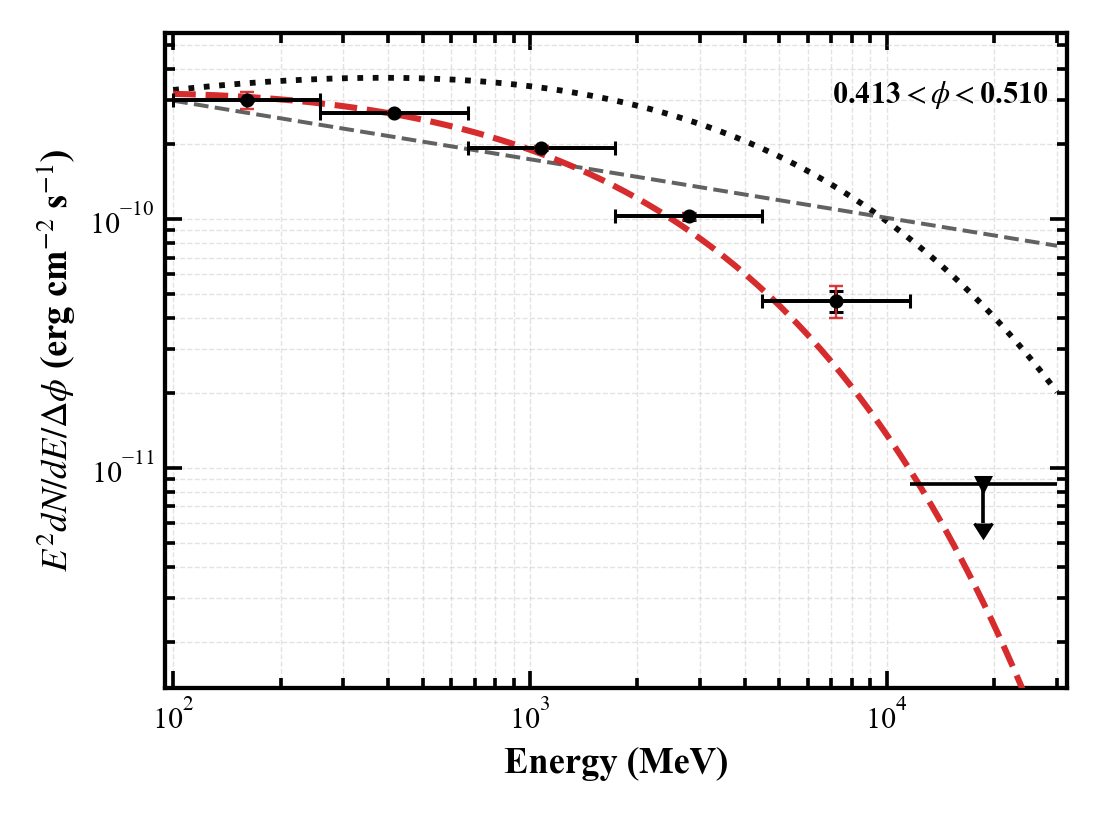}\hspace{0.012\textwidth}
	\includegraphics[width=0.42\textwidth]{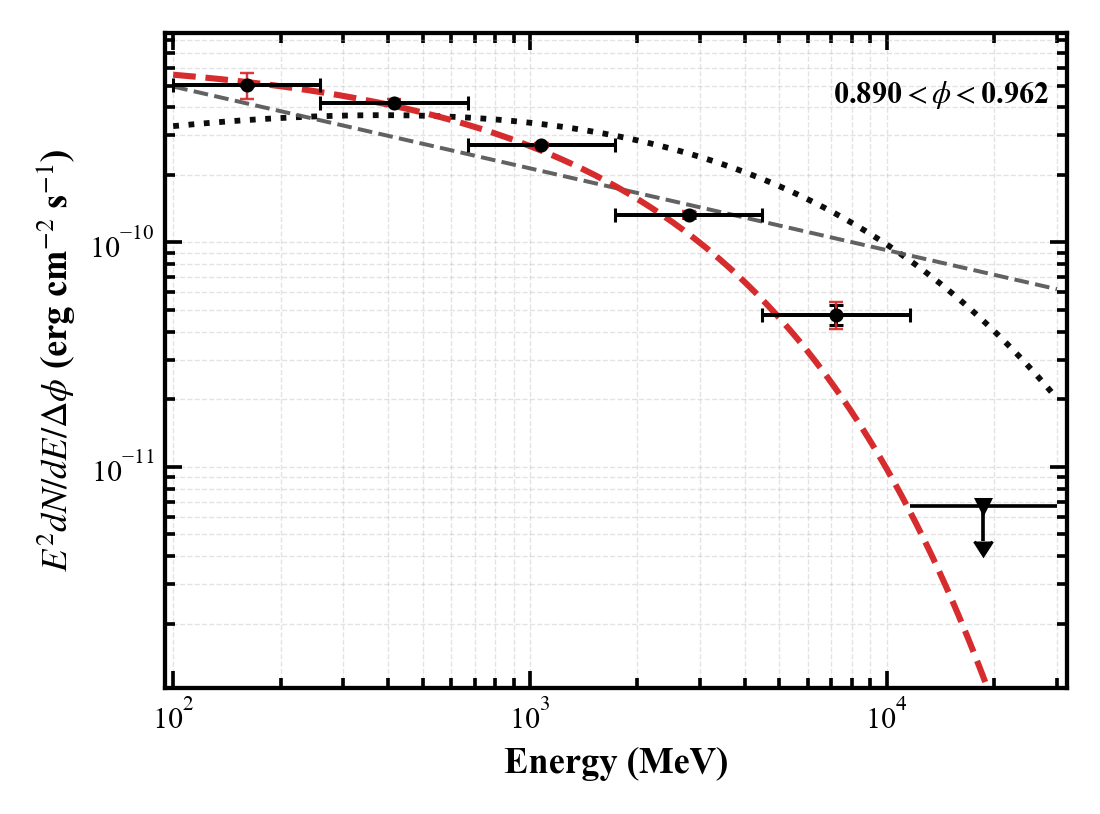}\\[-4pt]
	\includegraphics[width=0.42\textwidth]{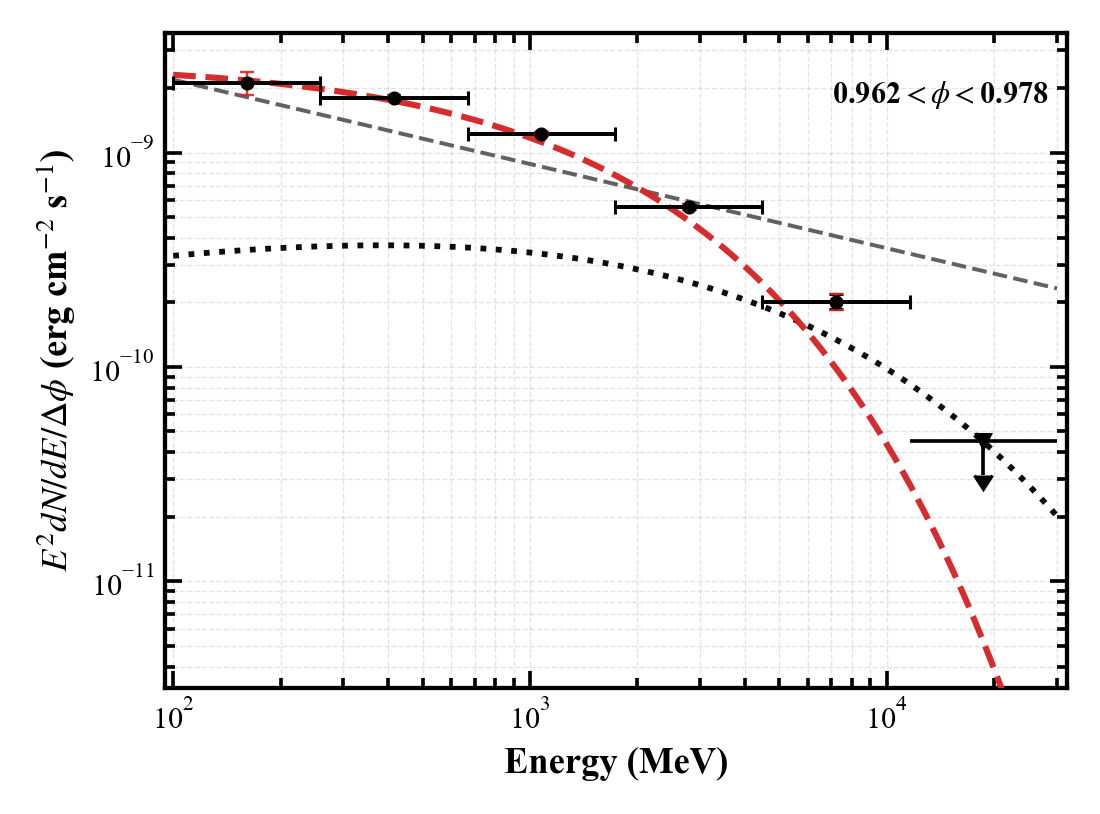}\hspace{0.012\textwidth}
	\includegraphics[width=0.42\textwidth]{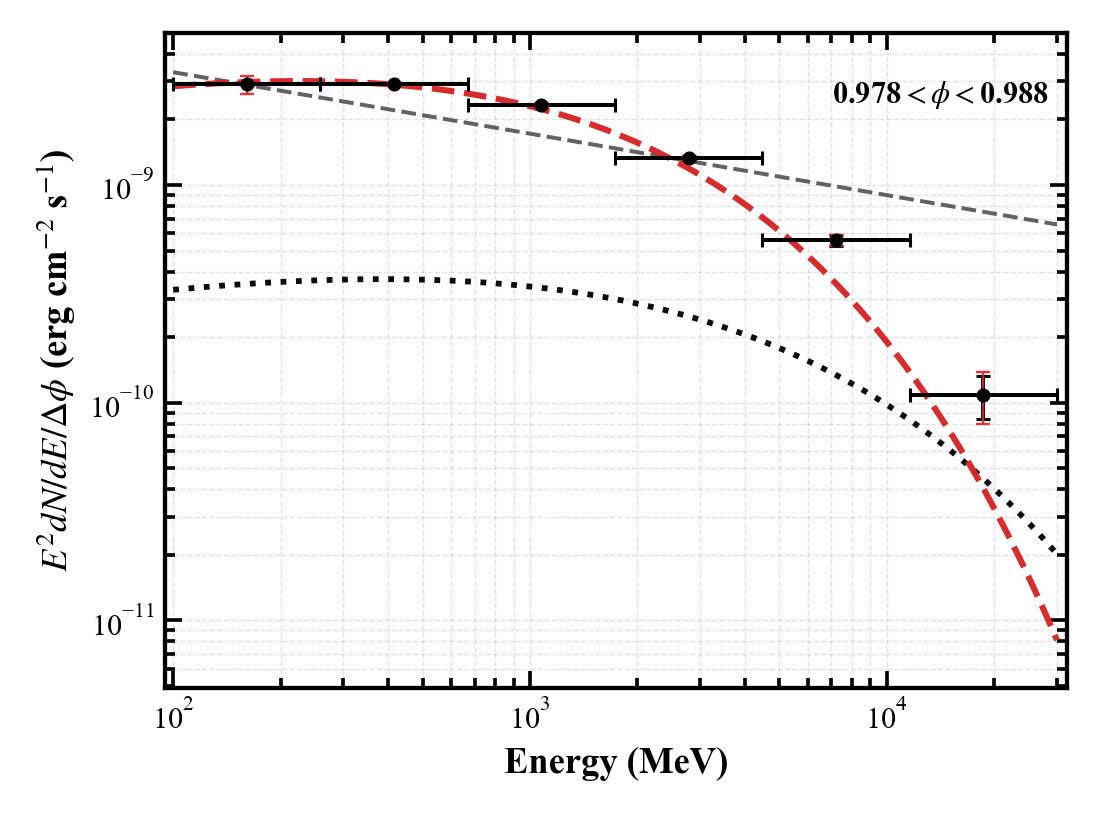}\\[-4pt]
	\includegraphics[width=0.42\textwidth]{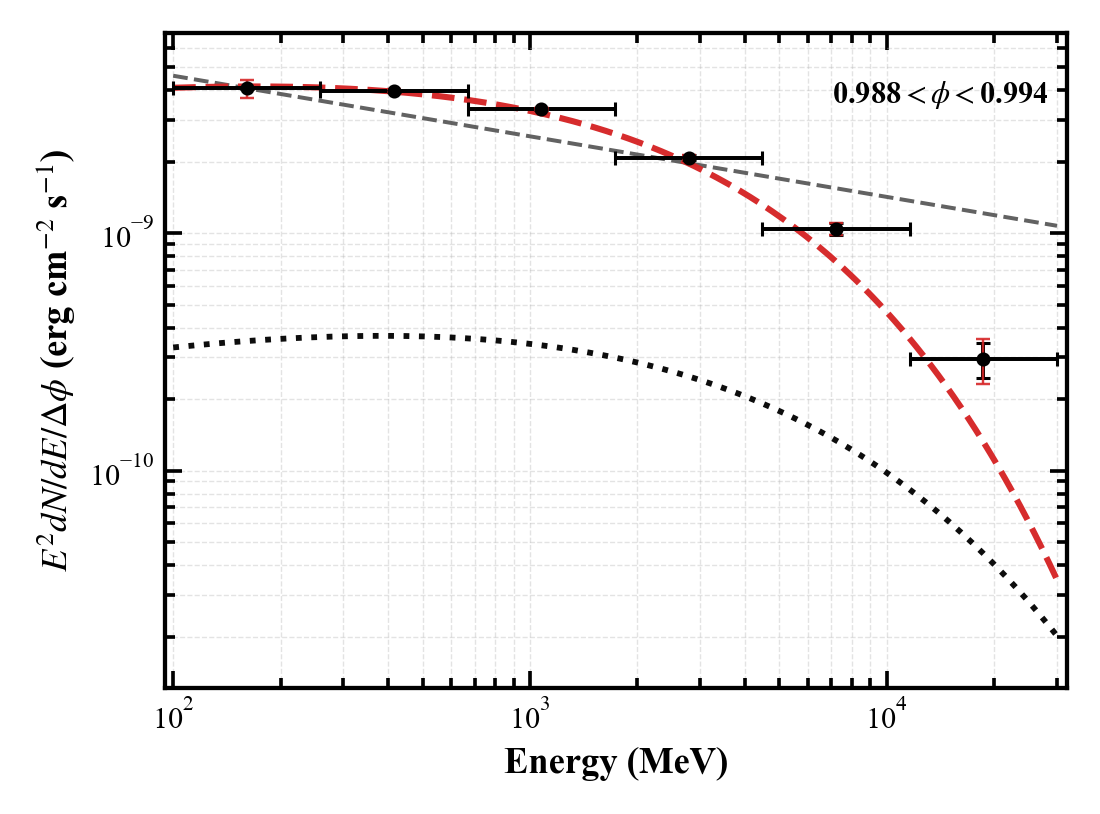}\hspace{0.012\textwidth}
	\includegraphics[width=0.42\textwidth]{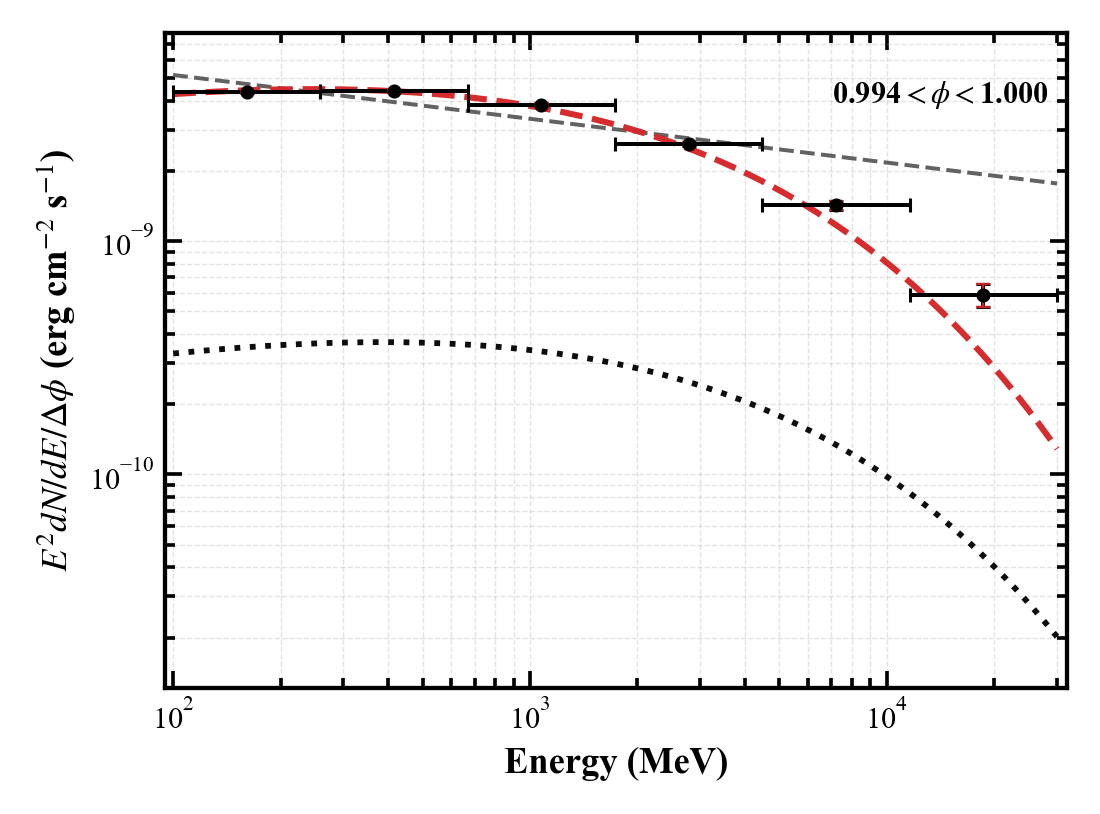}
	{%
		\addtocounter{figure}{-1}%
		\makeatletter\def\theHfigure{\thefigure.2}\makeatother%
		\caption{(Continued)}%
	}
\end{figure*}

\begin{table*}[t]
	\centering
	\scriptsize
	\setlength{\tabcolsep}{8pt}
	\caption{Phase-Resolved Spectral Parameters and Model Preference\label{tab:phase_params_placeholder}}
	\begin{tabular}{ccccccccc}
		\hline
		$\phi_{\min}$ & $\phi_{\max}$ & Flux$^{a}$ & $\Gamma_{\rm PL}$ & $\gamma_0$ & $d$ & $b$ & Preferred model & $\Delta\mathrm{AIC}$ \\
		\hline
		0.000 & 0.007 & $20.04\pm0.11$ & $2.17\pm0.01$ & $2.30\pm0.02$ & $0.28\pm0.01$ & $0.58\pm0.10$ & PLEC4 & 575.9 \\
		0.007 & 0.023 & $9.43\pm0.06$  & $2.19\pm0.01$ & $2.30\pm0.02$ & $0.28\pm0.02$ & $0.52\pm0.11$ & PLEC4 & 186.4 \\
		0.023 & 0.098 & $1.90\pm0.02$  & $2.07\pm0.01$ & $2.12\pm0.02$ & $0.13\pm0.02$ & $0.99\pm0.21$ & PL    & -95.7 \\
		0.098 & 0.289 & $0.88\pm0.01$  & $2.02\pm0.01$ & $2.07\pm0.02$ & $0.14\pm0.02$ & $0.97\pm0.20$ & PL    & -320.6 \\
		0.289 & 0.344 & $2.95\pm0.02$  & $2.10\pm0.01$ & $2.13\pm0.01$ & $0.24\pm0.02$ & $1.33\pm0.00$ & PLEC4 & 111.0 \\
		0.344 & 0.371 & $4.97\pm0.14$  & $2.15\pm0.01$ & $2.26\pm0.02$ & $0.31\pm0.01$ & $0.33\pm0.00$ & PLEC4 & 176.9 \\
		0.371 & 0.392 & $6.54\pm0.26$  & $2.14\pm0.01$ & $2.21\pm0.02$ & $0.26\pm0.01$ & $0.42\pm0.12$ & PLEC4 & 183.9 \\
		0.392 & 0.413 & $6.72\pm0.14$  & $2.16\pm0.01$ & $2.28\pm0.02$ & $0.24\pm0.01$ & $0.33\pm0.00$ & PLEC4 & 134.3 \\
		0.413 & 0.510 & $1.63\pm0.04$  & $2.24\pm0.01$ & $2.62\pm0.03$ & $0.41\pm0.05$ & $0.52\pm0.18$ & PL    & -18.9 \\
		0.890 & 0.962 & $2.56\pm0.02$  & $2.36\pm0.01$ & $2.75\pm0.03$ & $0.49\pm0.04$ & $0.65\pm0.14$ & PLEC4 & 118.8 \\
		0.962 & 0.978 & $10.77\pm0.06$ & $2.39\pm0.01$ & $2.74\pm0.03$ & $0.50\pm0.00$ & $0.63\pm0.08$ & PLEC4 & 300.6 \\
		0.978 & 0.988 & $16.36\pm1.85$ & $2.28\pm0.01$ & $2.53\pm0.02$ & $0.45\pm0.02$ & $0.50\pm0.09$ & PLEC4 & 488.6 \\
		0.988 & 0.994 & $23.24\pm0.50$ & $2.26\pm0.01$ & $2.41\pm0.02$ & $0.33\pm0.02$ & $0.56\pm0.10$ & PLEC4 & 303.4 \\
		0.994 & 1.000 & $26.21\pm0.14$ & $2.19\pm0.01$ & $2.34\pm0.02$ & $0.28\pm0.01$ & $0.43\pm0.09$ & PLEC4 & 579.3 \\
		\hline
		0.000 & 1.000 & $2.10\pm0.01$ & $2.10\pm0.00$ & $2.23\pm0.00$ & $0.23\pm0.00$ & $0.56\pm0.03$ & PLEC4 & 2415.5 \\
		\hline
	\end{tabular}
	\begin{minipage}{0.98\textwidth}
		\vspace{2pt}
		\footnotesize
			\textit{Note.} $^{a}$ Photon flux in 100 MeV$-$30 GeV divided by the phase-interval width, in units of \(10^{-6}\,\mathrm{ph\,cm^{-2}\,s^{-1}}\). Uncertainties are \(1\sigma\) statistical errors. The last row corresponds to the full phase interval (0.000$-$1.000). \(\Delta\mathrm{AIC}\equiv \mathrm{AIC}_{\rm PL}-\mathrm{AIC}_{\rm PLEC4}\), so positive values favor PLEC4.
	\end{minipage}
\end{table*}

\subsection{Energy-dependent Fixed-window Fractional Fluxes}
\begin{figure}
	\centering
	\includegraphics[width=0.48\textwidth]{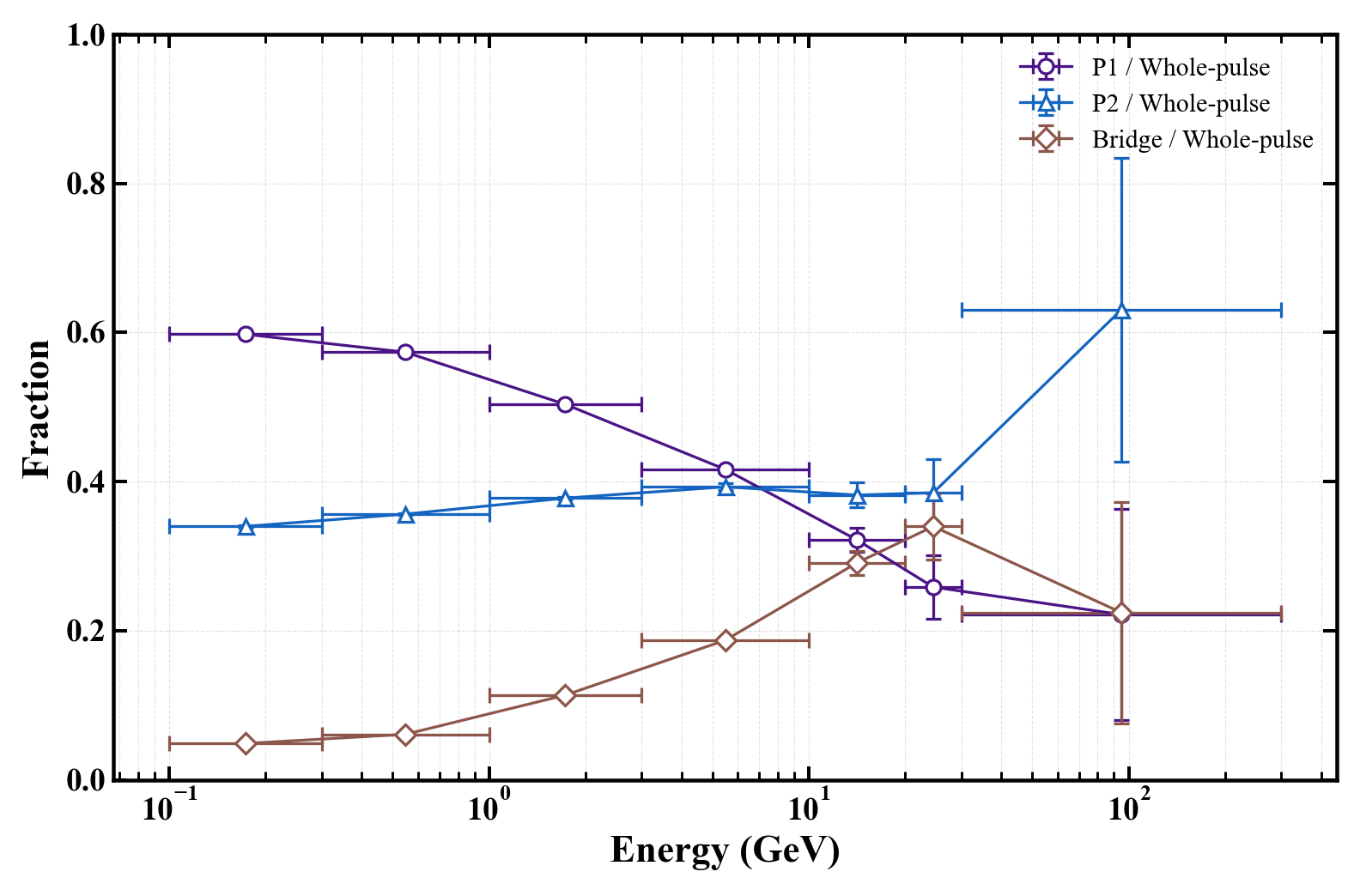}
		\caption{Energy dependence of the fixed-window fractional fluxes of the Crab pulsar in the same seven energy bands as Figure~\ref{fig:profiles_main}. Each point gives the OFF-subtracted pulsed counts in the fixed P1, P2, or Bridge window divided by the OFF-subtracted pulsed counts in the full phase interval. Vertical error bars show \(1\sigma\) statistical uncertainties propagated with the shared OFF term, and horizontal error bars indicate the widths of the energy bands.}
	\label{fig:phase_fraction_vs_energy}
\end{figure}

We next examined how the relative contributions of the fixed P1, P2, and Bridge windows evolve with energy. For this purpose, we measured their OFF-subtracted pulsed-count fractions relative to the OFF-subtracted pulsed counts in the full phase interval. Using the same phase-aligned FT1 dataset and the same energy-dependent angular cut as in Section~\ref{subsubsec:energy_resolved_profiles}, we computed the fixed-window pulsed counts using the same background-subtracted prescription as in Eq.~(2):
\begin{equation}
N_{\mathrm{pulsed},k}=N_k-\rho_{\mathrm{bkg}}\Delta\phi_k=N_k-\alpha_kN_{\rm off},
\end{equation}
where \(k\in\{P1,P2,\mathrm{Bridge},\mathrm{Full}\}\), \(\rho_{\mathrm{bkg}}=N_{\rm off}/\Delta\phi_{\rm off}\), and \(\alpha_k=\Delta\phi_k/\Delta\phi_{\rm off}\). For the full phase interval, \(\Delta\phi_{\rm Full}=1\), so \(\alpha_{\rm Full}=1/\Delta\phi_{\rm off}\). We then defined
\begin{equation}
f_k=\frac{N_{\mathrm{pulsed},k}}{N_{\mathrm{pulsed},\mathrm{Full}}}, \qquad k\in\{P1,P2,\mathrm{Bridge}\}.
\end{equation}
The statistical uncertainties were obtained by propagating the Poisson errors of the source-window, full-phase, and OFF counts, including the shared OFF term in both numerator and denominator. The resulting fractional fluxes are listed in Table~\ref{tab:phase_fraction} and shown in Figure~\ref{fig:phase_fraction_vs_energy}. At low energies, the pulsed emission is dominated by P1, with \(f_{P1}\approx 0.60\) in 0.1$-$0.3 GeV, compared with \(f_{P2}\approx 0.34\) and \(f_{\rm Bridge}\approx 0.05\). As the energy increases, the P1 fraction decreases steadily, while the P2 and Bridge fractions increase, and P1 and P2 become comparable at a few GeV. In the 10$-$20 and 20$-$30 GeV bands, \(f_{P2}\) exceeds \(f_{P1}\). In the 30$-$300 GeV band, the uncertainties become large, consistent with the absence of a significant pulse detection in that band. These fractional fluxes are consistent with the previously reported decline of the \(P1/P2\) ratio with energy \citep{abdo_fermi_2010,yeung_inferring_2020} and also show more clearly that the Bridge component becomes increasingly important toward higher energies.

\begin{table}[t]
\centering
\scriptsize
\setlength{\tabcolsep}{3.5pt}
\caption{Energy-dependent Fixed-window Fractional Fluxes\label{tab:phase_fraction}}
\begin{tabular*}{\columnwidth}{@{\extracolsep{\fill}}cccc@{}}
\hline
\shortstack{Energy Band\\(GeV)} & \shortstack{P1/\\Full phase} & \shortstack{P2/\\Full phase} & \shortstack{Bridge/\\Full phase} \\
\hline
0.1$-$0.3  & \(0.598\pm0.002\) & \(0.340\pm0.001\) & \(0.050\pm0.001\) \\
0.3$-$1    & \(0.574\pm0.001\) & \(0.357\pm0.001\) & \(0.062\pm0.001\) \\
1$-$3      & \(0.504\pm0.002\) & \(0.378\pm0.002\) & \(0.114\pm0.002\) \\
3$-$10     & \(0.417\pm0.004\) & \(0.393\pm0.004\) & \(0.188\pm0.004\) \\
10$-$20    & \(0.322\pm0.016\) & \(0.382\pm0.017\) & \(0.292\pm0.016\) \\
20$-$30    & \(0.259\pm0.042\) & \(0.385\pm0.045\) & \(0.340\pm0.044\) \\
30$-$300   & \(0.222\pm0.141\) & \(0.631\pm0.203\) & \(0.225\pm0.148\) \\
\hline
\end{tabular*}
\tablecomments{0.95\columnwidth}{Fractional fluxes are computed from the OFF-subtracted pulsed counts in the fixed P1, P2, and Bridge windows, each normalized by the OFF-subtracted pulsed counts in the full phase interval. Uncertainties are \(1\sigma\) statistical errors propagated with the shared OFF term.}
\end{table}

\section{Off-Pulse Analysis of the Crab Nebula}
\label{sec:offpulse_pwn}
\subsection{Off-Pulse LAT Data Preparation and Source Modeling}
\label{subsec:offpulse_data_model}
We used the off-pulse interval defined in Section~\ref{subsec:pulse_profiles} (0.51$-$0.89 in phase) to suppress the pulsed component and examine the baseline emission from the Crab Nebula region. Using the same procedure described in Section~\ref{sec:data_phase_assignment}, we constructed the phase-aligned FT1 event sample used for this off-pulse analysis. We then analyzed the off-pulse spectrum over 100 MeV$-$1 TeV using the same phase-aligned dataset. The likelihood exposure was scaled by the off-pulse phase width (\(\Delta\phi = 0.38\)).

For source modeling in the Crab region, we followed the 4FGL source model \citep{abdollahi_fermi_2020}. Because the residual emission in the off-pulse interval is dominated by the surrounding nebula, we modeled this emission with the two catalog sources 4FGL J0534.5+2201i (inverse-Compton component; hereafter IC) and 4FGL J0534.5+2201s (synchrotron component), and removed the catalog source associated with the pulsed emission (4FGL J0534.5+2200). To keep the off-pulse analysis consistent with the phase-resolved pulsar analysis, we used the same IRF and diffuse-model setup as in Section~\ref{sec:phase_resolved}, namely the P8R3\_SOURCE\_V3 IRF, with energy dispersion enabled for all source components except the Galactic and isotropic diffuse backgrounds, together with the \texttt{gll\_iem\_v07.fits} Galactic diffuse model and the corresponding P8R3\_SOURCE\_V3 isotropic template. In this fit, the spatial parameters of the nebular components, including the catalog extension of 4FGL J0534.5+2201i, were fixed. We did not attempt a new extension measurement in this paper.

We first ran an off-pulse fit over the 100 MeV$-$1 TeV energy range with a wide ROI (10$^\circ$ width, 0.04$^\circ$ pixel size, \texttt{zmax}=90, and \texttt{evtype}=3) to stabilize the local source model. Before fitting, the synchrotron component was reparameterized from the catalog LogParabola form to a point-source PowerLaw model. We allowed the normalization and index of the synchrotron component, the LogParabola parameters of the IC component (\(N\), \(\alpha\), \(\beta\)), and the Galactic and isotropic diffuse normalizations to vary. The resulting ROI model was then used as the starting point for a higher-energy refinement. We next restricted the fit to 1$-$1000 GeV, reduced the ROI to 6$^\circ$ with 0.025$^\circ$ pixels, and performed a joint likelihood fit over PSF0$-$PSF3 event types. In this step, the synchrotron component was fixed, whereas the IC spectral parameters and the diffuse normalizations were refit. This configuration was adopted for the subsequent off-pulse analysis.

\subsection{Off-Pulse Spectrum}
\label{subsec:offpulse_spectrum}
\begin{figure}
	\centering
	\includegraphics[width=0.48\textwidth]{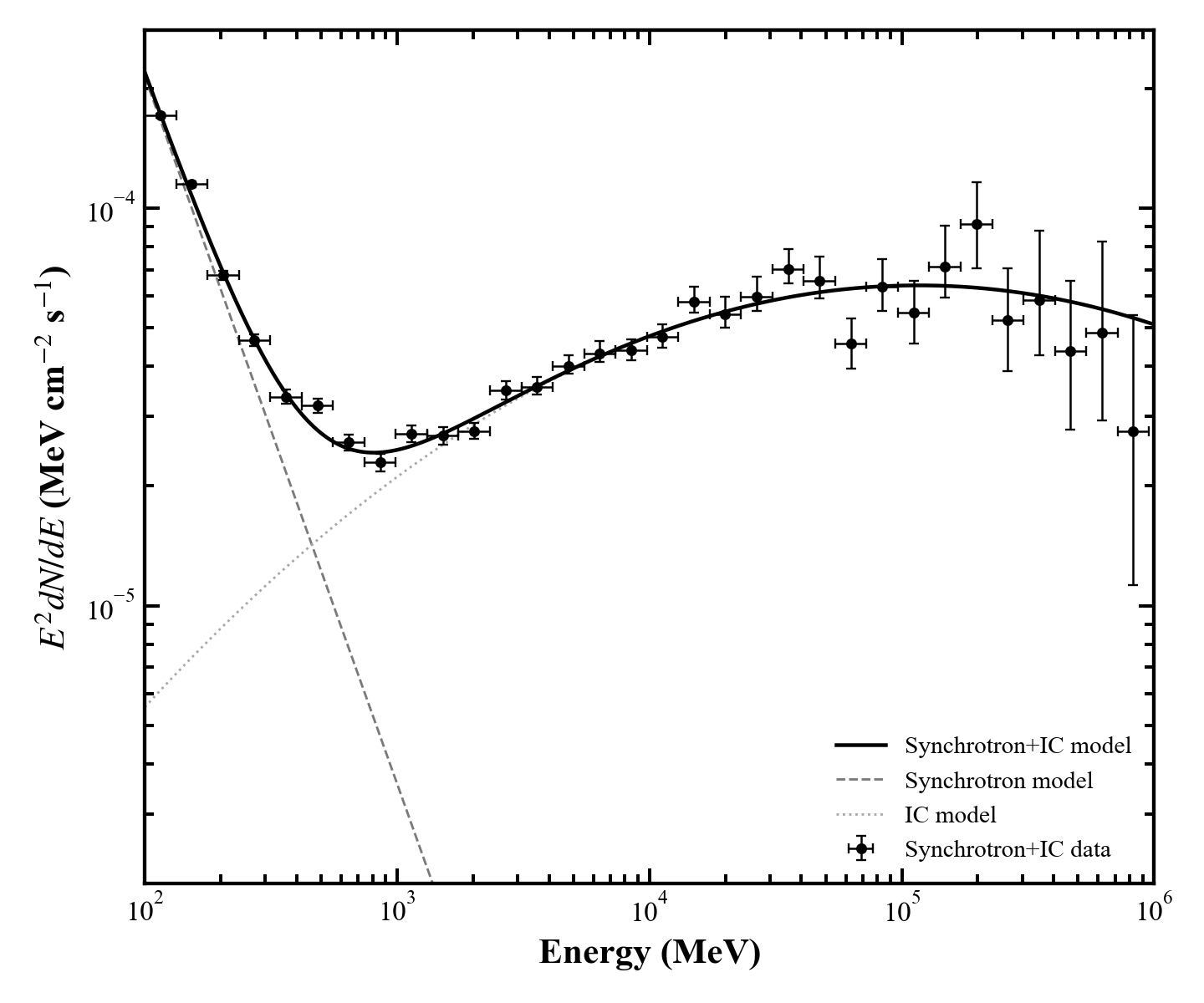}
		\caption{Off-pulse SED of the Crab nebula. The black points show a single set of spectral flux points for the combined Synchrotron+IC spectrum from the common-ROI bin-by-bin likelihood analysis. The gray dashed and gray dotted curves show the fitted synchrotron and IC component models, respectively, and the solid black curve gives their summed Synchrotron+IC model. Error bars indicate statistical uncertainties on the SED points.}
	\label{fig:pwn_sed}
\end{figure}

Figure~\ref{fig:pwn_sed} shows the combined Synchrotron+IC SED derived from the off-pulse interval. This emission is dominated by the synchrotron and IC components of the Crab nebula. We therefore derived a single set of spectral flux points for the combined Synchrotron+IC spectrum in the common ROI, rather than separate SED point sets for the two catalog components. The synchrotron-component fitted parameters were taken from the 100 MeV$-$1 TeV fit in Section~\ref{subsec:offpulse_data_model}, whereas the IC-component fitted parameters were taken from the subsequent PSF-partitioned refinement. The test statistic is defined as \(TS \equiv 2(\ln L_1-\ln L_0)\), where \(\ln L_0\) and \(\ln L_1\) are the log-likelihood values without and with the tested source component, respectively \citep{1996ApJ...461..396M}. Using the PowerLaw and LogParabola forms adopted in the LAT catalog framework \citep{abdollahi_fermi_2020,ballet_fermi_2024}, we obtained \(TS\sim 15725\) for the synchrotron component, with a PowerLaw index of \(3.77\pm0.04\) and differential normalization \(N_0=(3.60\pm0.30)\times10^{-12}\ \mathrm{MeV^{-1}\,cm^{-2}\,s^{-1}}\) at 1 GeV. The IC component yielded \(TS\sim 31570\) in the 100 MeV$-$1 TeV fit and \(TS\sim 28568\) in the PSF-partitioned refinement. With the pivot energy fixed at \(E_b=10\) GeV, the fitted LogParabola parameters were the differential normalization \(N=(4.81\pm0.10)\times10^{-13}\ \mathrm{MeV^{-1}\,cm^{-2}\,s^{-1}}\), the local spectral slope \(\alpha = 1.76\pm0.01\) at \(E_b\), and the curvature parameter \(\beta = 0.05\pm0.01\). This result is consistent with previous LAT nebular studies. \citet{2012ApJ...749...26B} provided an early benchmark for the average Crab Nebula spectrum using the first 33 months of Fermi observations and modeled it as the sum of a soft synchrotron component and a curved IC component. \citet{2020A&A...638A.147Y} analyzed 10 years of off-pulse LAT data over 60 MeV$-$10 GeV. They modeled the synchrotron and IC components with PL components and derived the SEDs for the total Synchrotron+IC emission. Their time-averaged spectral properties show \(\Gamma\simeq3.28\) for the synchrotron component and \(\Gamma\simeq1.42\) for the IC component. In the present off-pulse analysis of the Crab Nebula, we use the full 17-year LAT data set accumulated so far and extend the energy range to 1 TeV. The resulting nebular fit is consistent with previous off-pulse LAT analyses.

\section{Summary and Discussion}
\label{sec:summary_discussion}

In this work, we describe in detail the data preparation, processing, and analysis workflow for the \textit{Fermi}-LAT data analysis of the Crab pulsar. Using 17 years of \textit{Fermi}-LAT data, from 2008 August to 2025 August, we present a phase-aligned dataset of the Crab pulsar that extends the time span of earlier LAT studies. The larger photon statistics enable updated analyses of the pulse profiles and phase-resolved spectra, together with an off-pulse analysis of the same dataset. Pulsed emission remains clearly visible in the 10$-$20 and 20$-$30~GeV bands, allowing a more detailed characterization of the Crab pulsar at high energies. The fixed-window fractional fluxes show that the contribution of P1 decreases steadily with energy, while those of P2 and the Bridge increase, with P2 exceeding P1 above 10 GeV. The off-pulse analysis also confirms the synchrotron and IC components that dominate the emission in the selected off-pulse interval. Together, these results provide a detailed LAT-based view of the phase morphology, phase-resolved spectra, and off-pulse nebular emission of the Crab system.

In the future, as additional LAT data accumulate, the phase-aligned framework used here should further improve constraints on the pulse morphology and phase-resolved spectra of the Crab pulsar. In combination with multiwavelength and very-high-energy observations, it may also help clarify the origin of the high-energy emission and the connection between the pulsed and nebular components.

\begin{acknowledgements}
	
This work was partially supported by the National Natural Science Foundation of China (12233006, 12563010, 12503016), by Yunnan Fundamental Research Projects (grant No. 202501AS070068), and by the Project of Yunnan Provincial Department of Education Science Research Fund (2026Y0173).
This research made use of publicly available \textit{Fermi}-LAT data obtained from the FSSC data server and distributed by NASA Goddard Space Flight Center (GSFC).
\end{acknowledgements}

\bibliography{ms2026-0292}

@article{the_veritas_collaboration_detection_2011,
  author = {{The VERITAS Collaboration} and Aliu, E. and Arlen, T. and Aune, T. and Beilicke, M. and Benbow, W. and Bouvier, A. and Bradbury, S. M. and Buckley, J. H. and Bugaev, V. and Byrum, K. and Cannon, A. and Cesarini, A. and Christiansen, J. L. and Ciupik, L. and Collins-Hughes, E. and Connolly, M. P. and Cui, W. and Dickherber, R. and Duke, C. and Errando, M. and Falcone, A. and Finley, J. P. and Finnegan, G. and Fortson, L. and Furniss, A. and Galante, N. and Gall, D. and Gibbs, K. and Gillanders, G. H. and Godambe, S. and Griffin, S. and Grube, J. and Guenette, R. and Gyuk, G. and Hanna, D. and Holder, J. and Huan, H. and Hughes, G. and Hui, C. M. and Humensky, T. B. and Imran, A. and Kaaret, P. and Karlsson, N. and Kertzman, M. and Kieda, D. and Krawczynski, H. and Krennrich, F. and Lang, M. J. and Lyutikov, M. and Madhavan, A. S and Maier, G. and Majumdar, P. and McArthur, S. and McCann, A. and McCutcheon, M. and Moriarty, P. and Mukherjee, R. and Nuñez, P. and Ong, R. A. and Orr, M. and Otte, A. N. and Park, N. and Perkins, J. S. and Pizlo, F. and Pohl, M. and Prokoph, H. and Quinn, J. and Ragan, K. and Reyes, L. C. and Reynolds, P. T. and Roache, E. and Rose, H. J. and Ruppel, J. and Saxon, D. B. and Schroedter, M. and Sembroski, G. H. and Şentürk, G. D. and Smith, A. W. and Staszak, D. and Tešić, G. and Theiling, M. and Thibadeau, S. and Tsurusaki, K. and Tyler, J. and Varlotta, A. and Vassiliev, V. V. and Vincent, S. and Vivier, M. and Wakely, S. P. and Ward, J. E. and Weekes, T. C. and Weinstein, A. and Weisgarber, T. and Williams, D. A. and Zitzer, B.},
  title = {Detection of {Pulsed} {Gamma} {Rays} {Above} 100 {GeV} from the {Crab} {Pulsar}},
  journal = {Science},
  year = {2011},
  month = oct,
  volume = {334},
  number = {6052},
  pages = {69--72},
  doi = {10.1126/science.1208192},
  url = {https://www.science.org/doi/10.1126/science.1208192}
}

@article{aleksic_phase-resolved_2012,
  author = {Aleksić, J. and Alvarez, E. A. and Antonelli, L. A. and Antoranz, P. and Asensio, M. and Backes, M. and Barrio, J. A. and Bastieri, D. and Becerra González, J. and Bednarek, W. and Berdyugin, A. and Berger, K. and Bernardini, E. and Biland, A. and Blanch, O. and Bock, R. K. and Boller, A. and Bonnoli, G. and Borla Tridon, D. and Braun, I. and Bretz, T. and Cañellas, A. and Carmona, E. and Carosi, A. and Colin, P. and Colombo, E. and Contreras, J. L. and Cortina, J. and Cossio, L. and Covino, S. and Dazzi, F. and De Angelis, A. and De Caneva, G. and De Cea Del Pozo, E. and De Lotto, B. and Delgado Mendez, C. and Diago Ortega, A. and Doert, M. and Domínguez, A. and Dominis Prester, D. and Dorner, D. and Doro, M. and Eisenacher, D. and Elsaesser, D. and Ferenc, D. and Fonseca, M. V. and Font, L. and Fruck, C. and García López, R. J. and Garczarczyk, M. and Garrido, D. and Giavitto, G. and Godinović, N. and Hadasch, D. and Häfner, D. and Herrero, A. and Hildebrand, D. and Höhne-Mönch, D. and Hose, J. and Hrupec, D. and Jogler, T. and Kellermann, H. and Klepser, S. and Krähenbühl, T. and Krause, J. and Kushida, J. and La Barbera, A. and Lelas, D. and Leonardo, E. and Lewandowska, N. and Lindfors, E. and Lombardi, S. and López, M. and López-Oramas, A. and Lorenz, E. and Makariev, M. and Maneva, G. and Mankuzhiyil, N. and Mannheim, K. and Maraschi, L. and Mariotti, M. and Martínez, M. and Mazin, D. and Meucci, M. and Miranda, J. M. and Mirzoyan, R. and Moldón, J. and Moralejo, A. and Munar-Adrover, P. and Niedzwiecki, A. and Nieto, D. and Nilsson, K. and Nowak, N. and Orito, R. and Paneque, D. and Paoletti, R. and Pardo, S. and Paredes, J. M. and Partini, S. and Perez-Torres, M. A. and Persic, M. and Peruzzo, L. and Pilia, M. and Pochon, J. and Prada, F. and Prada Moroni, P. G. and Prandini, E. and Puerto Gimenez, I. and Puljak, I. and Reichardt, I. and Reinthal, R. and Rhode, W. and Ribó, M. and Rico, J. and Rügamer, S. and Saggion, A. and Saito, K. and Saito, T. Y. and Salvati, M. and Satalecka, K. and Scalzotto, V. and Scapin, V. and Schultz, C. and Schweizer, T. and Shayduk, M. and Shore, S. N. and Sillanpää, A. and Sitarek, J. and Šnidarić, I. and Sobczynska, D. and Spanier, F. and Spiro, S. and Stamatescu, V. and Stamerra, A. and Steinke, B. and Storz, J. and Strah, N. and Surić, T. and Takalo, L. and Takami, H. and Tavecchio, F. and Temnikov, P. and Terzić, T. and Tescaro, D. and Teshima, M. and Tibolla, O. and Torres, D. F. and Treves, A. and Uellenbeck, M. and Vankov, H. and Vogler, P. and Wagner, R. M. and Weitzel, Q. and Zabalza, V. and Zandanel, F. and Zanin, R. and Hirotani, K.},
  title = {Phase-resolved energy spectra of the {Crab} pulsar in the range of 50–400 {GeV} measured with the {MAGIC} telescopes},
  journal = {Astronomy \& Astrophysics},
  year = {2012},
  month = apr,
  volume = {540},
  pages = {A69},
  doi = {10.1051/0004-6361/201118166},
  url = {http://www.aanda.org/10.1051/0004-6361/201118166}
}

@article{ansoldi_teraelectronvolt_2016,
  author = {Ansoldi, S. and Antonelli, L. A. and Antoranz, P. and Babic, A. and Bangale, P. and Barres De Almeida, U. and Barrio, J. A. and Becerra González, J. and Bednarek, W. and Bernardini, E. and Biasuzzi, B. and Biland, A. and Blanch, O. and Bonnefoy, S. and Bonnoli, G. and Borracci, F. and Bretz, T. and Carmona, E. and Carosi, A. and Colin, P. and Colombo, E. and Contreras, J. L. and Cortina, J. and Covino, S. and Da Vela, P. and Dazzi, F. and De Angelis, A. and De Caneva, G. and De Lotto, B. and De Oña Wilhelmi, E. and Delgado Mendez, C. and Di Pierro, F. and Dominis Prester, D. and Dorner, D. and Doro, M. and Einecke, S. and Eisenacher Glawion, D. and Elsaesser, D. and Fernández-Barral, A. and Fidalgo, D. and Fonseca, M. V. and Font, L. and Frantzen, K. and Fruck, C. and Galindo, D. and García López, R. J. and Garczarczyk, M. and Garrido Terrats, D. and Gaug, M. and Godinović, N. and González Muñoz, A. and Gozzini, S. R. and Hanabata, Y. and Hayashida, M. and Herrera, J. and Hirotani, K. and Hose, J. and Hrupec, D. and Hughes, G. and Idec, W. and Kellermann, H. and Knoetig, M. L. and Kodani, K. and Konno, Y. and Krause, J. and Kubo, H. and Kushida, J. and La Barbera, A. and Lelas, D. and Lewandowska, N. and Lindfors, E. and Lombardi, S. and Longo, F. and López, M. and López-Coto, R. and López-Oramas, A. and Lorenz, E. and Makariev, M. and Mallot, K. and Maneva, G. and Mannheim, K. and Maraschi, L. and Marcote, B. and Mariotti, M. and Martínez, M. and Mazin, D. and Menzel, U. and Miranda, J. M. and Mirzoyan, R. and Moralejo, A. and Munar-Adrover, P. and Nakajima, D. and Neustroev, V. and Niedzwiecki, A. and Nevas Rosillo, M. and Nilsson, K. and Nishijima, K. and Noda, K. and Orito, R. and Overkemping, A. and Paiano, S. and Palatiello, M. and Paneque, D. and Paoletti, R. and Paredes, J. M. and Paredes-Fortuny, X. and Persic, M. and Poutanen, J. and Prada Moroni, P. G. and Prandini, E. and Puljak, I. and Reinthal, R. and Rhode, W. and Ribó, M. and Rico, J. and Rodriguez Garcia, J. and Saito, T. and Saito, K. and Satalecka, K. and Scalzotto, V. and Scapin, V. and Schultz, C. and Schweizer, T. and Shore, S. N. and Sillanpää, A. and Sitarek, J. and Snidaric, I. and Sobczynska, D. and Stamerra, A. and Steinbring, T. and Strzys, M. and Takalo, L. and Takami, H. and Tavecchio, F. and Temnikov, P. and Terzić, T. and Tescaro, D. and Teshima, M. and Thaele, J. and Torres, D. F. and Toyama, T. and Treves, A. and Ward, J. and Will, M. and Zanin, R.},
  title = {Teraelectronvolt pulsed emission from the {Crab} {Pulsar} detected by {MAGIC}},
  journal = {Astronomy \& Astrophysics},
  year = {2016},
  month = jan,
  volume = {585},
  pages = {A133},
  doi = {10.1051/0004-6361/201526853},
  url = {http://www.aanda.org/10.1051/0004-6361/201526853}
}

@ARTICLE{1999ApJ...516..297T,
  author = {{Thompson}, D.~J. and {Bailes}, M. and {Bertsch}, D.~L. and {Cordes}, J. and {D'Amico}, N. and {Esposito}, J.~A. and {Finley}, J. and {Hartman}, R.~C. and {Hermsen}, W. and {Kanbach}, G. and {Kaspi}, V.~M. and {Kniffen}, D.~A. and {Kuiper}, L. and {Lin}, Y.~C. and {Lyne}, A. and {Manchester}, R. and {Matz}, S.~M. and {Mayer-Hasselwander}, H.~A. and {Michelson}, P.~F. and {Nolan}, P.~L. and {{\"O}gelman}, H. and {Pohl}, M. and {Ramanamurthy}, P.~V. and {Sreekumar}, P. and {Reimer}, O. and {Taylor}, J.~H. and {Ulmer}, M.},
  title = "{Gamma Radiation from PSR B1055-52}",
  journal = {\apj},
  year = 1999,
  month = may,
  volume = {516},
  number = {1},
  pages = {297-306},
  doi = {10.1086/307083},
  adsurl = {https://ui.adsabs.harvard.edu/abs/1999ApJ...516..297T}
}

@article{abdo_fermi_2010,
  author = {Abdo, A. A. and Ackermann, M. and Ajello, M. and Atwood, W. B. and Axelsson, M. and Baldini, L. and Ballet, J. and Barbiellini, G. and Baring, M. G. and Bastieri, D. and Bechtol, K. and Bellazzini, R. and Berenji, B. and Blandford, R. D. and Bloom, E. D. and Bonamente, E. and Borgland, A. W. and Bregeon, J. and Brez, A. and Brigida, M. and Bruel, P. and Burnett, T. H. and Caliandro, G. A. and Cameron, R. A. and Camilo, F. and Caraveo, P. A. and Casandjian, J. M. and Cecchi, C. and Çelik, Ö. and Chekhtman, A. and Cheung, C. C. and Chiang, J. and Ciprini, S. and Claus, R. and Cognard, I. and Cohen-Tanugi, J. and Cominsky, L. R. and Conrad, J. and Dermer, C. D. and De Angelis, A. and De Luca, A. and De Palma, F. and Digel, S. W. and Do Couto E Silva, E. and Drell, P. S. and Dubois, R. and Dumora, D. and Espinoza, C. and Farnier, C. and Favuzzi, C. and Fegan, S. J. and Ferrara, E. C. and Focke, W. B. and Frailis, M. and Freire, P. C. C. and Fukazawa, Y. and Funk, S. and Fusco, P. and Gargano, F. and Gasparrini, D. and Gehrels, N. and Germani, S. and Giavitto, G. and Giebels, B. and Giglietto, N. and Giordano, F. and Glanzman, T. and Godfrey, G. and Grenier, I. A. and Grondin, M.-H. and Grove, J. E. and Guillemot, L. and Guiriec, S. and Hanabata, Y. and Harding, A. K. and Hayashida, M. and Hays, E. and Hughes, R. E. and Jóhannesson, G. and Johnson, A. S. and Johnson, R. P. and Johnson, T. J. and Johnson, W. N. and Johnston, S. and Kamae, T. and Katagiri, H. and Kataoka, J. and Kawai, N. and Kerr, M. and Knödlseder, J. and Kocian, M. L. and Kramer, M. and Kuehn, F. and Kuss, M. and Lande, J. and Latronico, L. and Lee, S.-H. and Lemoine-Goumard, M. and Longo, F. and Loparco, F. and Lott, B. and Lovellette, M. N. and Lubrano, P. and Lyne, A. G. and Makeev, A. and Marelli, M. and Mazziotta, M. N. and McEnery, J. E. and Meurer, C. and Michelson, P. F. and Mitthumsiri, W. and Mizuno, T. and Moiseev, A. A. and Monte, C. and Monzani, M. E. and Moretti, E. and Morselli, A. and Moskalenko, I. V. and Murgia, S. and Nakamori, T. and Nolan, P. L. and Norris, J. P. and Noutsos, A. and Nuss, E. and Ohsugi, T. and Omodei, N. and Orlando, E. and Ormes, J. F. and Ozaki, M. and Paneque, D. and Panetta, J. H. and Parent, D. and Pelassa, V. and Pepe, M. and Pesce-Rollins, M. and Pierbattista, M. and Piron, F. and Porter, T. A. and Rainò, S. and Rando, R. and Ray, P. S. and Razzano, M. and Reimer, A. and Reimer, O. and Reposeur, T. and Ritz, S. and Rochester, L. S. and Rodriguez, A. Y. and Romani, R. W. and Roth, M. and Ryde, F. and Sadrozinski, H. F.-W. and Sanchez, D. and Sander, A. and Parkinson, P. M. Saz and Scargle, J. D. and Sgrò, C. and Siskind, E. J. and Smith, D. A. and Smith, P. D. and Spandre, G. and Spinelli, P. and Stappers, B. W. and Strickman, M. S. and Suson, D. J. and Tajima, H. and Takahashi, H. and Tanaka, T. and Thayer, J. B. and Thayer, J. G. and Theureau, G. and Thompson, D. J. and Thorsett, S. E. and Tibaldo, L. and Torres, D. F. and Tosti, G. and Tramacere, A. and Uchiyama, Y. and Usher, T. L. and Van Etten, A. and Vasileiou, V. and Vilchez, N. and Vitale, V. and Waite, A. P. and Wallace, E. and Wang, P. and Watters, K. and Weltevrede, P. and Winer, B. L. and Wood, K. S. and Ylinen, T. and Ziegler, M.},
  title = {\textit{{FERMI}} {LARGE} {AREA} {TELESCOPE} {OBSERVATIONS} {OF} {THE} {CRAB} {PULSAR} {AND} {NEBULA}},
  journal = {The Astrophysical Journal},
  year = {2010},
  month = jan,
  volume = {708},
  number = {2},
  pages = {1254--1267},
  doi = {10.1088/0004-637X/708/2/1254},
  url = {https://iopscience.iop.org/article/10.1088/0004-637X/708/2/1254}
}

@article{kerr_timing_2015,
  author = {Kerr, M. and Ray, P. S. and Johnston, S. and Shannon, R. M. and Camilo, F.},
  title = {{TIMING} {GAMMA}-{RAY} {PULSARS} {WITH} {THE} \textit{{FERMI}} {LARGE} {AREA} {TELESCOPE}: {TIMING} {NOISE} {AND} {ASTROMETRY}},
  journal = {The Astrophysical Journal},
  year = {2015},
  month = nov,
  volume = {814},
  number = {2},
  pages = {128},
  doi = {10.1088/0004-637X/814/2/128},
  url = {https://iopscience.iop.org/article/10.1088/0004-637X/814/2/128}
}

@article{pshirkov_gamma-ray_2020,
  author = {Pshirkov, M S and Nizamov, B A and Bykov, A M and Uvarov, Yu A},
  title = {Gamma-ray flux depressions of the {Crab} {Nebula} in the high-energy range},
  journal = {Monthly Notices of the Royal Astronomical Society},
  year = {2020},
  month = aug,
  volume = {496},
  number = {4},
  pages = {5227--5232},
  doi = {10.1093/mnras/staa1921},
  url = {https://academic.oup.com/mnras/article/496/4/5227/5872499}
}

@ARTICLE{1993MNRAS.265.1003L,
  author = {{Lyne}, A.~G. and {Pritchard}, R.~S. and {Graham Smith}, F.},
  title = "{23 years of Crab pulsar rotational history.}",
  journal = {\mnras},
  year = 1993,
  month = dec,
  volume = {265},
  pages = {1003-1012},
  doi = {10.1093/mnras/265.4.1003},
  adsurl = {https://ui.adsabs.harvard.edu/abs/1993MNRAS.265.1003L}
}

@article{yeung_inferring_2020,
  author = {Yeung, Paul K. H.},
  title = {Inferring the origins of the pulsed \textit{γ} -ray emission from the {Crab} pulsar with ten-year \textit{{Fermi}} -{LAT} data},
  journal = {Astronomy \& Astrophysics},
  year = {2020},
  month = aug,
  volume = {640},
  pages = {A43},
  doi = {10.1051/0004-6361/202038166},
  url = {https://www.aanda.org/10.1051/0004-6361/202038166}
}

@ARTICLE{2024A&A...690A.167A,
  author = {{Abe}, K. and {Abe}, S. and {Abhishek}, A. and {Acero}, F. and {Aguasca-Cabot}, A. and {Agudo}, I. and {Alvarez Crespo}, N. and {Antonelli}, L.~A. and {Aramo}, C. and {Arbet-Engels}, A. and {Arcaro}, C. and {Artero}, M. and {Asano}, K. and {Aubert}, P. and {Baktash}, A. and {Bamba}, A. and {Baquero Larriva}, A. and {Baroncelli}, L. and {Barres de Almeida}, U. and {Barrio}, J.~A. and {Batkovic}, I. and {Baxter}, J. and {Becerra Gonz{\'a}ilez}, J. and {Bernardini}, E. and {Bernete Medrano}, J. and {Berti}, A. and {Bhattacharjee}, P. and {Bigongiari}, C. and {Bissaldi}, E. and {Blanch}, O. and {Bonnoli}, G. and {Bordas}, P. and {Brunelli}, G. and {Bulgarelli}, A. and {Burelli}, I. and {Burmistrov}, L. and {Buscemi}, M. and {Cardillo}, M. and {Caroff}, S. and {Carosi}, A. and {Carrasco}, M.~S. and {Cassol}, F. and {Castrej{\'o}n}, N. and {Cauz}, D. and {Cerasole}, D. and {Ceribella}, G. and {Chai}, Y. and {Cheng}, K. and {Chiavassa}, A. and {Chikawa}, M. and {Chon}, G. and {Chytka}, L. and {Cicciari}, G.~M. and {Cifuentes}, A. and {Contreras}, J.~L. and {Cortina}, J. and {Costantini}, H. and {Da Vela}, P. and {Dalchenko}, M. and {Dazzi}, F. and {De Angelis}, A. and {de Bony de Lavergne}, M. and {De Lotto}, B. and {de Menezes}, R. and {Del Peral}, L. and {Delgado}, C. and {Delgado Mengual}, J. and {della Volpe}, D. and {Dellaiera}, M. and {Di Piano}, A. and {Di Pierro}, F. and {Di Tria}, R. and {Di Venere}, L. and {D{\'\i}az}, C. and {Dominik}, R.~M. and {Dominis Prester}, D. and {Donini}, A. and {Dorner}, D. and {Doro}, M. and {Eisenberger}, L. and {Els{\"a}sser}, D. and {Emery}, G. and {Escudero}, J. and {Fallah Ramazani}, V. and {Ferrarotto}, F. and {Fiasson}, A. and {Foffano}, L. and {Freixas Coromina}, L. and {Fr{\"o}se}, S. and {Fukazawa}, Y. and {Garcia L{\'o}pez}, R. and {Gasbarra}, C. and {Gasparrini}, D. and {Gavira}, L. and {Geyer}, D. and {Giesbrecht Paiva}, J. and {Giglietto}, N. and {Giordano}, F. and {Gliwny}, P. and {Godinovic}, N. and {Grau}, R. and {Green}, D. and {Green}, J. and {Gunji}, S. and {G{\"u}nther}, P. and {Hackfeld}, J. and {Hadasch}, D. and {Hahn}, A. and {Hassan}, T. and {Hayashi}, K. and {Heckmann}, L. and {Heller}, M. and {Herrera Llorente}, J. and {Hirotani}, K. and {Hoffmann}, D. and {Horns}, D. and {Houles}, J. and {Hrabovsky}, M. and {Hrupec}, D. and {Hui}, D. and {Iarlori}, M. and {Imazawa}, R. and {Inada}, T. and {Inome}, Y. and {Ioka}, K. and {Iori}, M. and {Jimenez Martinez}, I. and {Jim{\'e}nez Quiles}, J. and {Jurysek}, J. and {Kagaya}, M. and {Karas}, V. and {Katagiri}, H. and {Kataoka}, J. and {Kerszberg}, D. and {Kobayashi}, Y. and {Kohri}, K. and {Kong}, A. and {Kubo}, H. and {Kushida}, J. and {Lainez}, M. and {Lamanna}, G. and {Lamastra}, A. and {Lemoigne}, L. and {Linhoff}, M. and {Longo}, F. and {L{\'o}pez-Coto}, R. and {L{\'o}pez-Moya}, M. and {L{\'o}pez-Oramas}, A. and {Loporchio}, S. and {Lorini}, A. and {Lozano Bahilo}, J. and {Luque-Escamilla}, P.~L. and {Majumdar}, P. and {Makariev}, M. and {Mallamaci}, M. and {Mandat}, D. and {Manganaro}, M. and {Manic{\`o}}, G. and {Mannheim}, K. and {Marchesi}, S. and {Mariotti}, M. and {Marquez}, P. and {Marsella}, G. and {Mart{\'\i}}, J. and {Martinez}, O. and {Mart{\'\i}nez}, G. and {Mart{\'\i}nez}, M. and {Mas-Aguilar}, A. and {Maurin}, G. and {Mazin}, D. and {Mestre Guillen}, E. and {Micanovic}, S. and {Miceli}, D. and {Miener}, T. and {Miranda}, J.~M. and {Mirzoyan}, R. and {Mizuno}, T. and {Molero Gonzalez}, M. and {Molina}, E. and {Montaruli}, T. and {Moralejo}, A. and {Morcuende}, D. and {Morselli}, A. and {Moya}, V. and {Muraishi}, H. and {Nagataki}, S. and {Nakamori}, T. and {Neronov}, A. and {Nickel}, L. and {Nievas Rosillo}, M. and {Nikolic}, L. and {Nishijima}, K. and {Noda}, K. and {Nosek}, D. and {Novotny}, V. and {Nozaki}, S. and {Ohishi}, M. and {Ohtani}, Y. and {Oka}, T. and {Okumura}, A.},
  title = "{A detailed study of the very high-energy Crab pulsar emission with the LST-1}",
  journal = {\aap},
  year = 2024,
  month = oct,
  volume = {690},
  eid = {A167},
  pages = {A167},
  doi = {10.1051/0004-6361/202450059},
  adsurl = {https://ui.adsabs.harvard.edu/abs/2024A&A...690A.167A}
}

@misc{ballet_fermi_2024,
  author = {Ballet, J. and Bruel, P. and Burnett, T. H. and Lott, B. and collaboration, The Fermi-LAT},
  title = {Fermi {Large} {Area} {Telescope} {Fourth} {Source} {Catalog} {Data} {Release} 4 ({4FGL}-{DR4})},
  year = {2024},
  month = jul,
  doi = {10.48550/arXiv.2307.12546},
  url = {http://arxiv.org/abs/2307.12546},
  publisher = {arXiv},
  note = {arXiv:2307.12546 [astro-ph]}
}

@article{meyer_characterizing_2019,
  author = {Meyer, Manuel and Scargle, Jeffrey D. and Blandford, Roger D.},
  title = {Characterizing the {Gamma}-{Ray} {Variability} of the {Brightest} {Flat} {Spectrum} {Radio} {Quasars} {Observed} with the {Fermi} {LAT}},
  journal = {The Astrophysical Journal},
  year = {2019},
  month = may,
  volume = {877},
  number = {1},
  pages = {39},
  doi = {10.3847/1538-4357/ab1651},
  url = {https://iopscience.iop.org/article/10.3847/1538-4357/ab1651}
}

@article{aharonian_spectrum_2024,
  author = {Aharonian, F. and Benkhali, F. Ait and Aschersleben, J. and Ashkar, H. and Backes, M. and Baktash, A. and Martins, V. Barbosa and Batzofin, R. and Becherini, Y. and Berge, D. and Bernlöhr, K. and Bi, B. and Böttcher, M. and Boisson, C. and Bolmont, J. and Lavergne, M. de Bony de and Borowska, J. and Bradascio, F. and Breuhaus, M. and Brose, R. and Brown, A. and Brun, F. and Bruno, B. and Bulik, T. and Burger-Scheidlin, C. and Bylund, T. and Caroff, S. and Casanova, S. and Cecil, R. and Celic, J. and Cerruti, M. and Chambery, P. and Chand, T. and Chandra, S. and Chen, A. and Chibueze, J. and Chibueze, O. and Cotter, G. and Cristofari, P. and Devin, J. and Djannati-Ataï, A. and Djuvsland, J. and Dmytriiev, A. and Einecke, S. and Ernenwein, J.-P. and Fegan, S. and Feijen, K. and Filipović, M. and Fontaine, G. and Füßling, M. and Funk, S. and Gabici, S. and Gallant, Y. A. and Giavitto, G. and Glawion, D. and Glicenstein, J. F. and Glombitza, J. and Goswami, P. and Grolleron, G. and Grondin, M.-H. and Haerer, L. and Hinton, J. A. and Hofmann, W. and Holch, T. L. and Holler, M. and Horns, D. and Jamrozy, M. and Jankowsky, F. and Joshi, V. and Kasai, E. and Katarzyński, K. and Khatoon, R. and Khélifi, B. and Kluźniak, W. and Komin, Nu and Kosack, K. and Kostunin, D. and Kundu, A. and Lang, R. G. and Stum, S. Le and Leitl, F. and Lemière, A. and Lemoine-Goumard, M. and Lenain, J.-P. and Leuschner, F. and Luashvili, A. and Mackey, J. and Malyshev, D. and Malyshev, D. and Marandon, V. and Marinos, P. and Martí-Devesa, G. and Marx, R. and Mehta, A. and Meyer, M. and Mitchell, A. and Moderski, R. and Mohrmann, L. and Montanari, A. and Moulin, E. and Murach, T. and Naurois, M. de and Niemiec, J. and O'Brien, P. and Ohm, S. and Olivera-Nieto, L. and Wilhelmi, E. de Ona and Ostrowski, M. and Panny, S. and Panter, M. and Parsons, R. D. and Peron, G. and Prokhorov, D. A. and Pühlhofer, G. and Punch, M. and Quirrenbach, A. and Regeard, M. and Reichherzer, P. and Reimer, A. and Reimer, O. and Ren, H. and Renaud, M. and Reville, B. and Rieger, F. and Roellinghoff, G. and Rudak, B. and Sahakian, V. and Salzmann, H. and Sasaki, M. and Schüssler, F. and Schutte, H. M. and Shapopi, J. N. S. and Specovius, A. and Spencer, S. and Stawarz, Ł and Steenkamp, R. and Steinmassl, S. and Steppa, C. and Streil, K. and Sushch, I. and Suzuki, H. and Takahashi, T. and Tanaka, T. and Terrier, R. and Tluczykont, M. and Tsuji, N. and Unbehaun, T. and Eldik, C. van and Vecchi, M. and Veh, J. and Venter, C. and Vink, J. and Wach, T. and Wagner, S. J. and Wierzcholska, A. and Zacharias, M. and Zargaryan, D. and Zdziarski, A. A. and Zech, A. and Zouari, S. and Żywucka, N. and Harding, A.},
  title = {Spectrum and extension of the inverse-{Compton} emission of the {Crab} {Nebula} from a combined {Fermi}-{LAT} and {H}.{E}.{S}.{S}. analysis},
  journal = {Astronomy \& Astrophysics},
  year = {2024},
  month = jun,
  volume = {686},
  pages = {A308},
  doi = {10.1051/0004-6361/202348651}
}

@ARTICLE{2010A&A...517L...9D,
  author = {{de Jager}, O.~C. and {B{\"u}sching}, I.},
  title = "{The H-test probability distribution revisited: improved sensitivity}",
  journal = {\aap},
  year = 2010,
  month = jul,
  volume = {517},
  eid = {L9},
  pages = {L9},
  doi = {10.1051/0004-6361/201014362},
  adsurl = {https://ui.adsabs.harvard.edu/abs/2010A&A...517L...9D}
}

@ARTICLE{2021ApJ...911...45L,
  author = {{Luo}, Jing and {Ransom}, Scott and {Demorest}, Paul and {Ray}, Paul S. and {Archibald}, Anne and {Kerr}, Matthew and {Jennings}, Ross J. and {Bachetti}, Matteo and {van Haasteren}, Rutger and {Champagne}, Chloe A. and et al.},
  title = "{PINT: A Modern Software Package for Pulsar Timing}",
  journal = {\apj},
  year = 2021,
  month = apr,
  volume = {911},
  number = {1},
  eid = {45},
  pages = {45},
  doi = {10.3847/1538-4357/abe62f},
  adsurl = {https://ui.adsabs.harvard.edu/abs/2021ApJ...911...45L}
}

@article{smith_third_2023,
  author = {Smith, David A. and Bruel, Philippe and Clark, Colin J. and Guillemot, Lucas and Kerr, Matthew T. and Ray, Paul and Abdollahi, Soheila and Ajello, Marco and Baldini, Luca and Ballet, Jean and Baring, Matthew and Bassa, Cees and Gonzalez, Josefa Becerra and Bellazzini, Ronaldo and Berretta, Alessandra and Bhattacharyya, Bhaswati and Bissaldi, Elisabetta and Bonino, Raffaella and Bottacini, Eugenio and Bregeon, Johan and Burgay, Marta and Burnett, Toby and Cameron, Rob and Camilo, Fernando and Caputo, Regina and Caraveo, Patrizia and Cavazzuti, Elisabetta and Chiaro, Graziano and Ciprini, Stefano and Cognard, Ismael and Orestano, Paolo Cristarella and Crnogorcevic, Milena and Cuoco, Alessandro and Cutini, Sara and D'Ammando, Filippo and Angelis, Alessandro de and Gaetano, Salvatore De and Menezes, Raniere de and Palma, Francesco de and DeCesar, Megan and Deneva, Julia and Lalla, Niccola Di and Venere, Leonardo Di and Dirirsa, Feraol Fana and Dominguez, Alberto and Dumora, Denis and Fegan, Stephen and Ferrara, Elizabeth and Fiori, Alessio and Fleischhack, Henrike and Flynn, Chris and Franckowiak, Anna and Freire, Paulo and Fukazawa, Yasushi and Fusco, Piergiorgio and Galanti, Giorgio and Gammaldi, Viviana and Gargano, Fabio and Gasparrini, Dario and Giacchino, Federica and Giglietto, Nico and Giordano, Francesco and Giroletti, Marcello and Green, David and Grenier, Isabelle and Guiriec, Sylvain and Gustafsson, Michael and Harding, Alice and Hays, Liz and Hewitt, John and Horan, Deirdre and Hou, Xian and Jankowski, Fabian and Johnson, Tyrel and Johnson, Robert and Johnston, Simon and Kataoka, Jun and Keith, Michael J. and Kramer, Michael and Kuss, Michael and Latronico, Luca and Lee, Shiu-Hang and Li, Di and Li, Jian and Limyansky, Brent and Longo, Francesco and Loparco, Francesco and Lorusso, Leonarda and Lovellette, Michael and Lower, Marcus and Lubrano, Pasquale and Lyne, Andrew and Maldera, Simone and Manchester, Richard and Manfreda, Alberto and Marelli, Martino and Marta-Devesa, Guillem and Mazziotta, Mario Nicola and McEnery, Julie and Mereu, Isabella and Michelson, Peter and Mitthumsiri, Warit and Mizuno, Tsunefumi and Moiseev, Alex and Monzani, Maria Elena and Morselli, Aldo and Negro, Michela and Nemmen, Rodrigo and Nieder, Lars and Nuss, Eric and Omodei, Nicola and Orienti, Monica and Orlando, Elena and Ormes, Jonathan F. and Palatiello, Michele and Paneque, David and Panzarini, Giuliana and Persic, Massimo and Pesce-Rollins, Melissa and Pillera, Roberta and Poon, Helen and Porter, Troy and Principe, Giacomo and Raino, Silvia and Rando, Riccardo and Ransom, Scott and Razzano, Massimiliano and Razzaque, Soebur and Reimer, Anita and Reimer, Olaf and Renault-Tinacci, Nicolas and Romani, Roger and Sanchez-Conde, Miguel A. and Parkinson, Pablo Saz and Scotton, Lorenzo and Serini, Davide and Sgro, Carmelo and Shannon, Ryan and Sharma, Vidushi and Siskind, Eric J. and Spandre, Gloria and Spinelli, Paolo and Stappers, Ben and Stephens, Tom and Suson, Dan and Tajima, Hiro and Tak, Dongguen and Theureau, Gilles and Thompson, David and Tibolla, Omar and Torres, Diego F. and Valverde, Janeth and Venter, Christo and Wadiasingh, Zorawar and Wang, Nina and Wang, Pei and Weltevrede, Patrick and Wood, Kent and Zaharijas, Gabrijela},
  title = {The {Third} {Fermi} {Large} {Area} {Telescope} {Catalog} of {Gamma}-ray {Pulsars}},
  journal = {The Astrophysical Journal},
  year = {2023},
  month = dec,
  volume = {958},
  number = {2},
  pages = {191},
  doi = {10.3847/1538-4357/acee67}
}

@misc{lange_fermiphased_2025,
  author = {Lange, Alexander and K, B. B.},
  title = {{FermiPhased}: {A} tool for phase-resolved likelihood analysis of {Fermi}-{LAT} data},
  year = {2025},
  month = nov,
  doi = {10.48550/arXiv.2511.04810},
  url = {http://arxiv.org/abs/2511.04810},
  publisher = {arXiv},
  note = {arXiv:2511.04810 [astro-ph]}
}

@article{abdollahi_incremental_2022,
  author = {Abdollahi, S. and Acero, F. and Baldini, L. and Ballet, J. and Bastieri, D. and Bellazzini, R. and Berenji, B. and Berretta, A. and Bissaldi, E. and Blandford, R. D. and Bloom, E. and Bonino, R. and Brill, A. and Britto, R. J. and Bruel, P. and Burnett, T. H. and Buson, S. and Cameron, R. A. and Caputo, R. and Caraveo, P. A. and Castro, D. and Chaty, S. and Cheung, C. C. and Chiaro, G. and Cibrario, N. and Ciprini, S. and Coronado-Blázquez, J. and Crnogorcevic, M. and Cutini, S. and D’Ammando, F. and De Gaetano, S. and Digel, S. W. and Di Lalla, N. and Dirirsa, F. and Di Venere, L. and Domínguez, A. and Fallah Ramazani, V. and Fegan, S. J. and Ferrara, E. C. and Fiori, A. and Fleischhack, H. and Franckowiak, A. and Fukazawa, Y. and Funk, S. and Fusco, P. and Galanti, G. and Gammaldi, V. and Gargano, F. and Garrappa, S. and Gasparrini, D. and Giacchino, F. and Giglietto, N. and Giordano, F. and Giroletti, M. and Glanzman, T. and Green, D. and Grenier, I. A. and Grondin, M.-H. and Guillemot, L. and Guiriec, S. and Gustafsson, M. and Harding, A. K. and Hays, E. and Hewitt, J. W. and Horan, D. and Hou, X. and Jóhannesson, G. and Karwin, C. and Kayanoki, T. and Kerr, M. and Kuss, M. and Landriu, D. and Larsson, S. and Latronico, L. and Lemoine-Goumard, M. and Li, J. and Liodakis, I. and Longo, F. and Loparco, F. and Lott, B. and Lubrano, P. and Maldera, S. and Malyshev, D. and Manfreda, A. and Martí-Devesa, G. and Mazziotta, M. N. and Mereu, I. and Meyer, M. and Michelson, P. F. and Mirabal, N. and Mitthumsiri, W. and Mizuno, T. and Moiseev, A. A. and Monzani, M. E. and Morselli, A. and Moskalenko, I. V. and Negro, M. and Nuss, E. and Omodei, N. and Orienti, M. and Orlando, E. and Paneque, D. and Pei, Z. and Perkins, J. S. and Persic, M. and Pesce-Rollins, M. and Petrosian, V. and Pillera, R. and Poon, H. and Porter, T. A. and Principe, G. and Rainò, S. and Rando, R. and Rani, B. and Razzano, M. and Razzaque, S. and Reimer, A. and Reimer, O. and Reposeur, T. and Sánchez-Conde, M. and Saz Parkinson, P. M. and Scotton, L. and Serini, D. and Sgrò, C. and Siskind, E. J. and Smith, D. A. and Spandre, G. and Spinelli, P. and Sueoka, K. and Suson, D. J. and Tajima, H. and Tak, D. and Thayer, J. B. and Thompson, D. J. and Torres, D. F. and Troja, E. and Valverde, J. and Wood, K. and Zaharijas, G.},
  title = {Incremental {Fermi} {Large} {Area} {Telescope} {Fourth} {Source} {Catalog}},
  journal = {The Astrophysical Journal Supplement Series},
  year = {2022},
  month = jun,
  volume = {260},
  number = {2},
  pages = {53},
  doi = {10.3847/1538-4365/ac6751},
  url = {https://iopscience.iop.org/article/10.3847/1538-4365/ac6751}
}

@ARTICLE{1974ITAC...19..716A,
  author = {{Akaike}, H.},
  title = "{A New Look at the Statistical Model Identification}",
  journal = {IEEE Transactions on Automatic Control},
  year = 1974,
  month = jan,
  volume = {19},
  pages = {716-723},
  doi = {10.1109/TAC.1974.1100705},
  adsurl = {https://ui.adsabs.harvard.edu/abs/1974ITAC...19..716A}
}

@article{wood_fermipy_nodate,
  author = {Wood, Matthew},
  title = {Fermipy {Documentation}},
  journal = {Fermipy Documentation},
  year = {2017}
}

@article{abdollahi_fermi_2020,
  author = {Abdollahi, S. and Acero, F. and Ackermann, M. and Ajello, M. and Atwood, W. B. and Axelsson, M. and Baldini, L. and Ballet, J. and Barbiellini, G. and Bastieri, D. and Becerra Gonzalez, J. and Bellazzini, R. and Berretta, A. and Bissaldi, E. and Blandford, R. D. and Bloom, E. D. and Bonino, R. and Bottacini, E. and Brandt, T. J. and Bregeon, J. and Bruel, P. and Buehler, R. and Burnett, T. H. and Buson, S. and Cameron, R. A. and Caputo, R. and Caraveo, P. A. and Casandjian, J. M. and Castro, D. and Cavazzuti, E. and Charles, E. and Chaty, S. and Chen, S. and Cheung, C. C. and Chiaro, G. and Ciprini, S. and Cohen-Tanugi, J. and Cominsky, L. R. and Coronado-Blázquez, J. and Costantin, D. and Cuoco, A. and Cutini, S. and D’Ammando, F. and DeKlotz, M. and Torre Luque, P. De La and De Palma, F. and Desai, A. and Digel, S. W. and Lalla, N. Di and Mauro, M. Di and Venere, L. Di and Domínguez, A. and Dumora, D. and Dirirsa, F. Fana and Fegan, S. J. and Ferrara, E. C. and Franckowiak, A. and Fukazawa, Y. and Funk, S. and Fusco, P. and Gargano, F. and Gasparrini, D. and Giglietto, N. and Giommi, P. and Giordano, F. and Giroletti, M. and Glanzman, T. and Green, D. and Grenier, I. A. and Griffin, S. and Grondin, M.-H. and Grove, J. E. and Guiriec, S. and Harding, A. K. and Hayashi, K. and Hays, E. and Hewitt, J. W. and Horan, D. and Jóhannesson, G. and Johnson, T. J. and Kamae, T. and Kerr, M. and Kocevski, D. and Kovac’evic’, M. and Kuss, M. and Landriu, D. and Larsson, S. and Latronico, L. and Lemoine-Goumard, M. and Li, J. and Liodakis, I. and Longo, F. and Loparco, F. and Lott, B. and Lovellette, M. N. and Lubrano, P. and Madejski, G. M. and Maldera, S. and Malyshev, D. and Manfreda, A. and Marchesini, E. J. and Marcotulli, L. and Martí-Devesa, G. and Martin, P. and Massaro, F. and Mazziotta, M. N. and McEnery, J. E. and Mereu, I. and Meyer, M. and Michelson, P. F. and Mirabal, N. and Mizuno, T. and Monzani, M. E. and Morselli, A. and Moskalenko, I. V. and Negro, M. and Nuss, E. and Ojha, R. and Omodei, N. and Orienti, M. and Orlando, E. and Ormes, J. F. and Palatiello, M. and Paliya, V. S. and Paneque, D. and Pei, Z. and Peña-Herazo, H. and Perkins, J. S. and Persic, M. and Pesce-Rollins, M. and Petrosian, V. and Petrov, L. and Piron, F. and Poon, H. and Porter, T. A. and Principe, G. and Rainò, S. and Rando, R. and Razzano, M. and Razzaque, S. and Reimer, A. and Reimer, O. and Remy, Q. and Reposeur, T. and Romani, R. W. and Parkinson, P. M. Saz and Schinzel, F. K. and Serini, D. and Sgrò, C. and Siskind, E. J. and Smith, D. A. and Spandre, G. and Spinelli, P. and Strong, A. W. and Suson, D. J. and Tajima, H. and Takahashi, M. N. and Tak, D. and Thayer, J. B. and Thompson, D. J. and Tibaldo, L. and Torres, D. F. and Torresi, E. and Valverde, J. and Klaveren, B. Van and Zyl, P. Van and Wood, K. and Yassine, M. and Zaharijas, G.},
  title = {Fermi {Large} {Area} {Telescope} {Fourth} {Source} {Catalog}},
  journal = {The Astrophysical Journal Supplement Series},
  year = {2020},
  month = mar,
  volume = {247},
  number = {1},
  pages = {33},
  doi = {10.3847/1538-4365/ab6bcb},
  url = {https://iopscience.iop.org/article/10.3847/1538-4365/ab6bcb}
}

@ARTICLE{1996ApJ...461..396M,
  author = {{Mattox}, J.~R. and {Bertsch}, D.~L. and {Chiang}, J. and {Dingus}, B.~L. and {Digel}, S.~W. and {Esposito}, J.~A. and {Fierro}, J.~M. and {Hartman}, R.~C. and {Hunter}, S.~D. and {Kanbach}, G. and et al.},
  title = "{The Likelihood Analysis of EGRET Data}",
  journal = {\apj},
  year = 1996,
  month = apr,
  volume = {461},
  pages = {396},
  doi = {10.1086/177068},
  adsurl = {https://ui.adsabs.harvard.edu/abs/1996ApJ...461..396M}
}

@ARTICLE{2012ApJ...749...26B,
  author = {{Buehler}, R. and {Scargle}, J.~D. and {Blandford}, R.~D. and {Baldini}, L. and {Baring}, M.~G. and {Belfiore}, A. and {Charles}, E. and {Chiang}, J. and {D'Ammando}, F. and {Dermer}, C.~D. and {Funk}, S. and {Grove}, J.~E. and {Harding}, A.~K. and {Hays}, E. and {Kerr}, M. and {Massaro}, F. and {Mazziotta}, M.~N. and {Romani}, R.~W. and {Saz Parkinson}, P.~M. and {Tennant}, A.~F. and {Weisskopf}, M.~C.},
  title = "{Gamma-Ray Activity in the Crab Nebula: The Exceptional Flare of 2011 April}",
  journal = {\apj},
  year = 2012,
  month = apr,
  volume = {749},
  number = {1},
  eid = {26},
  pages = {26},
  doi = {10.1088/0004-637X/749/1/26},
  adsurl = {https://ui.adsabs.harvard.edu/abs/2012ApJ...749...26B}
}

@ARTICLE{2020A&A...638A.147Y,
  author = {{Yeung}, Paul K.~H. and {Horns}, Dieter},
  title = "{Fermi Large Area Telescope observations of the fast-dimming Crab Nebula in 60-600 MeV}",
  journal = {\aap},
  year = 2020,
  month = jun,
  volume = {638},
  eid = {A147},
  pages = {A147},
  doi = {10.1051/0004-6361/201936740},
  adsurl = {https://ui.adsabs.harvard.edu/abs/2020A&A...638A.147Y}
}
\bibliographystyle{raa}

\end{document}